\documentclass[fleqn,usenatbib]{mnras}

\usepackage{newtxtext,newtxmath}

\usepackage[T1]{fontenc}

\DeclareRobustCommand{\VAN}[3]{#2}
\let\VANthebibliography\thebibliography
\def\thebibliography{\DeclareRobustCommand{\VAN}[3]{##3}\VANthebibliography}

\usepackage{graphicx}	
\usepackage{amsmath}	
\usepackage{amssymb}	

\title[Elemental depletion]{Influence of galactic arm scale dynamics on the molecular composition of the cold and dense ISM III. Elemental depletion and shortcomings of the current physico-chemical models}

\author[Wakelam et al.]{
V. Wakelam$^{1}$, W. Iqbal$^{2}$, J.-P., Melisse$^{1}$, P. Gratier$^{1}$, M. Ruaud$^{3}$, I. Bonnell$^4$ 
\thanks{E-mail: valentine.wakelam@u-bordeaux.fr}
\\
$^{1}$Laboratoire d'astrophysique de Bordeaux, Univ. Bordeaux, CNRS, B18N, all\'ee Geoffroy Saint-Hilaire, 33615 Pessac, France\\
$^2$ South-Western Institute for Astronomy Research (SWIFAR), Yunnan University (YNU), Kunming 650500, People's Republic of China\\
$^3$ NASA Ames Research Center, Moffett Field, CA, USA\\
$^4$ Scottish Universities Physics Alliance (SUPA), School of Physics and Astronomy, University of St. Andrews, North Haugh,
St Andrews, Fife KY16 9SS, UK
}

\date{Accepted 2020 July 7. Received 2020 July 7; in original form 2020 January 14.}

\pubyear{2019}

\begin{document}
\maketitle
\begin{abstract}
We present a study of the elemental depletion in the interstellar medium. We combined the results of a Galatic model describing the  gas physical conditions during the formation of dense cores with a full-gas-grain chemical model. During the transition between diffuse and dense medium, the reservoirs of elements, initially atomic in the gas, are gradually depleted on dust grains (with a phase of neutralisation for those which are ions). This process becomes efficient when the density is larger than 100~cm$^{-3}$. If the dense material goes back into diffuse conditions, these elements are brought back in the gas-phase because of photo-dissociations of the molecules on the ices followed by thermal desorption from the grains. Nothing remains on the grains for densities below 10~cm$^{-3}$ or in the gas-phase in a molecular form. One exception is chlorine, which is efficiently converted at low density. Our current gas-grain chemical model is not able to reproduce the depletion of atoms observed in the diffuse medium except for Cl which gas abundance follows the observed one in medium with densities smaller than 10~cm$^{-3}$. This is an indication that crucial processes (involving maybe chemisorption and/or ice irradiation profoundly modifying the nature of the ices) are missing.
\end{abstract}




\section{Introduction}

After the Big Bang, the very first stars formed only from hydrogen (H) and helium (He).  Heavier elements are formed later, either through fusion inside stars or by neutron capture processes in supernovae explosion or neutron star mergers. At the end of the life of a star, some of the ejected atoms form refractory dust, others remain in the gas phase. This material is included in the cycle of interstellar matter: forming denser regions up to molecular clouds, newly stars and planetary systems, until the star dies and spreads its inner material into the diffuse interstellar medium (ISM) again. Within this cycle, chemical elements can be found in three phases: (a) in the gas-phase, (b) in the refractory grain cores, and (c) in the icy mantles of grains. The sum of abundance of the elements contained in all three phases is called the cosmic elemental abundance. Cosmic abundances are assumed to represent some reference value and are measured in the atmosphere of stars (our Sun or other stars) \citep[see for instance][]{1994ApJ...430..650S, 2009ARA&A..47..481A}. \\
In the diffuse medium, no icy mantles are expected to be present on grains and so the measured abundances of gas-phase elements (which are mostly in the ionized form except for O, N, and F) subtracted from these cosmic abundances should give the amount of each element stored in the refractory cores, and so not available for any volatile chemistry. Observations of gas-phase atomic lines towards different lines of sight indicate that the depletion of the elements from the gas-phase increases with the density of the cloud, even when the density of the cloud remains too low to explain the depletion simply by collisions with grains \citep{1996ARA&A..34..279S,2009ApJ...700.1299J}. Silicon, for instance, has a cosmic elemental abundance of $3.5\times 10^{-5}$ \citep[compared to H,][]{2009ARA&A..47..481A}. The gas-phase abundance Si$^+$ measured in clouds with densities of about $10^{-2}$ cm$^{-3}$ is $2\times 10^{-5}$ and drops by a factor of ten at densities of 10 cm$^{-3}$. This observed decrease of the atomic abundance in the gas-phase could be related to a more efficient neutralisation of the cations (since what is measured is Si$^+$), but at such low densities, only dielectronic recombination or ion-molecule reactions could really alter the balance of ion stages as it is the case for a small number of elements \citep{1996ARA&A..34..279S}. Observations of atomic lines in diffuse clouds can only be done for densities below 10 cm$^{-3}$ because at higher density the atomic lines become optically thick. In dense molecular clouds (with densities of a few $10^4$ cm$^{-3}$), SiO was not detected \citep[upper limit of $\sim 2\times 10^{-12}$ compared to H,][]{1989ApJ...343..201Z}, indicating that the depletion of silicon continues at densities larger than 10 cm$^{-3}$.

The mechanism of depletion can simply be understood as a matter of collision between gas-phase species and grains. At a density of 10 cm$^{-3}$ for a gas temperature of 100 K, the typical adsorption time is about $10^8$ yr, which is of the same order as the typical lifetime of interstellar grains in the ISM \citep{2009ASPC..414..453D}. At lower density, the adsorption timescale may be longer and turbulent mixing may play a role in driving dust from denser regions \citep{1998ApJ...499..267T}. The chemistry associated with this depletion and the resulting species are still a matter of debate \citep[see for instance][]{1998ApJ...499..267T,2010ApJ...710.1009W}. \citet{2010ApJ...710.1009W} for instance debated on the depletion of oxygen at different densities. Summing all the observed volatile phases of oxygen and the oxygen presumably contained in the refractory parts of the grains, \citeauthor{2010ApJ...710.1009W} found that approximately 28\% of the oxygen could not be accounted for, calling this fraction the unidentified depleted oxygen. 
 This very specific problem of the elemental depletion in tenuous regions (with densities below a few $10^3$~cm$^{-3}$) is particularly important for the chemistry of dense regions such as star and planetary  system forming regions. In fact, the fraction of the elements available for phases (a - gas-phase) and (c - ice mantle), i.e. only phases that can be observed in star forming regions, is determined by what is depleted in phase (b - grain core), which is quite uncertain. Considering the uncertainties in the fraction of elements still available, various values are used in astrochemical models and sometime adjusted to reproduce observations. 

In this paper, we explore this problem coupling time dependent simulations of the gas physical conditions following the transition between diffuse and dense interstellar regions with a detailed gas-grain chemical model. In two previous papers, we have already analyzed theses simulations to show 1) the chemical diversity of cold cores induced by the variety of physical histories experienced by the gas and dust forming such structures \citep{2018A&A...611A..96R}, and 2) the impact of these histories in the interstellar O$_2$ abundance \citep{2019MNRAS.486.4198W}. 

\section{Models description}

To simulate the chemistry during the formation of cold cores, we used the physical structure provided by the 3D SPH galactic model from \citet{2013MNRAS.430.1790B}. In these simulations the gas from the spiral arm is cold ($\rm T_{gas} < 100$~K) and $\rm n_H\sim100$~cm$^{-3}$, whilst gas entering comprises warm gas ($\rm T_{gas} < 8000$~K) and some cooler and denser gas from previous spiral arm encounters \citep{2013MNRAS.430.1790B,2018A&A...611A..96R}. Dense clouds are formed by converging flows when the interstellar gas enters the spiral arms. The gas is shocked and compressed which heats the gas, balanced against the increased atomic cooling rates at the higher densities \citep[Bonnell et al. 2013, and see also ][]{2007ApJ...657..870V,2002ApJ...564L..97K}.  Ultimately the cooling dominates and results in cold dense gas susceptible to gravitational collapse \citep[see also ][]{2005MNRAS.359..211L}. \citet{2013MNRAS.430.1790B} presented three types of simulations with three different spatial resolutions from the scale of a spiral galaxy, to the formation of dense clouds in the ISM, to the relatively small, sub-pc scale where star formation occurs. To keep the computational time reasonable, only the highest resolution sub-pc scale simulations include self-gravity. For the purpose of this work (because we are not interested in the collapse phase and we want to study the effect of the dense gas being brought back into the diffuse medium), we have used the "medium" spatial resolution, which does not include self-gravity such that if the external pressure decreases, the cool gas can re-expand and warm up.
The SPH model gives us the gas temperature, and density as a function of time and position in space. Using this information, we computed visual extinctions and dust temperatures as explained in \citet{2018A&A...611A..96R}. At the end of the simulations, twelve clouds were identified with a maximum peak density above $10^5$~cm$^{-3}$. We then identified all the SPH particles in a sphere of 0.5 pc around this maximum and retrieved the past and future history of these particles. The physical conditions of each of the particles were used as input parameters at each time step of the Nautilus 3-phase gas-grain model \citep{2016MNRAS.459.3756R}.  
In total, we have approximately 3000 trajectories describing the physical and chemical evolution of parcels of material in a galactic arm going from very diffuse conditions (below 0.1~cm$^{-3}$) to dense ones (of a few $10^5$~cm$^{-3}$). The SPH model does not include self gravity so the clouds are always dissipated after reaching the maximum peak density. This allows us to study the effect of cycles between dense and less dense ISM, which were suggested by \citet{2009ASPC..414..453D} as an explanation for elemental depletion. While exposing the results, we will make a distinction between the first phase of the simulation (the increase of the density with time up to a maximum value) and a second phase which begins when the cloud starts to dissipate and the density decreases again. 
More details on the methodology can be found in \citet{2018A&A...611A..96R}. Details on the chemical model and the chemical parameters are in \citet{2019MNRAS.486.4198W}. Analysis of the results of the physical model is done in \citet{2013MNRAS.430.1790B} and Bonnell et al. (in prep). The cosmic-ray ionisation rate is kept constant to $10^{-17}$~s$^{-1}$ for simplicity. Determining the charge of the interstellar grains is complex and it depends on many parameters relative to the grains themselves (size, nature etc) and on the physical properties of the environment (UV field, density, temperature) \citep{1989A&A...208..331B}. To limit the computational time, the astrochemical model used in these simulations assumes only one single grain size and the grains can be either neutral or negatively charged. Only neutral gas-phase species are allowed to stick on the grains as the grains are supposed to be mostly negatively charged when the density is above $10^3$~cm$^{-3}$ \citep{1989A&A...208..331B}. We do not take into account the possible sticking of ions on neutral grains. This means that we may overestimate the timescale of depletion of atoms on the grains as it may occur earlier in the simulations. At the lowest densities however, the electronic recombination of ionized atoms should be faster than the collision with grains. \\
Table~\ref{ab_elem} lists the elemental abundances used in the model. These values correspond to abundances observed in the most diffuse regions in \citet{2009ApJ...700.1299J} sample where they observed the smallest depletion (their F$^*$ = 0). These values are smaller than the solar abundances as the fraction of elements included in the refractory grains produced by stars has already disappeared. Two elements are not included in  \citet{2009ApJ...700.1299J} study: Na and F. For these two ones, we have used the solar abundance from \citet{2009ARA&A..47..481A}.

\begin{table}
\caption{Elemental abundances (with respect to the total hydrogen density) used in the simulations. The values are from \citet{2009ApJ...700.1299J} for a depletion factor (F$^*$) of zero except for Na and F for which we used the solar abundances from \citet{2009ARA&A..47..481A}.\label{ab_elem}}
\begin{center}
\begin{tabular}{|c|c|}
\hline
\hline
Atoms & Abundance \\
\hline
He & $9\times 10^{-2}$\\ 
N   & $6.2\times 10^{-5}$\\ 
O   & $5.6\times 10^{-4}$\\ 
C  & $2.4\times 10^{-4}$\\ 
S   & $1.3\times 10^{-5}$\\ 
Si   & $2.4\times 10^{-5}$\\ 
Fe  & $3.9\times 10^{-6}$\\ 
Na  & $1.7\times 10^{-6}$\\ 
Mg  & $2.2\times 10^{-5}$\\ 
P  & $6.8\times 10^{-7}$\\ 
Cl  & $5.9\times 10^{-7}$\\ 
F   & $3.6\times 10^{-8}$\\ 
\hline
\end{tabular}
\end{center}
\label{ab_elem}
\end{table}%

\section{Results}

\begin{figure*}
 \includegraphics[width=0.3\linewidth]{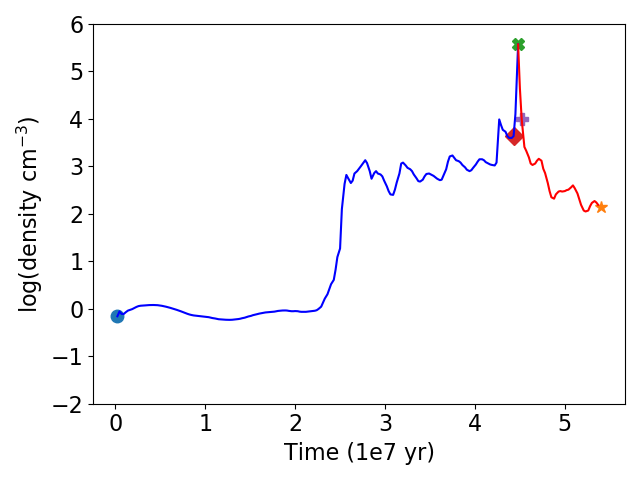}
 \includegraphics[width=0.3\linewidth]{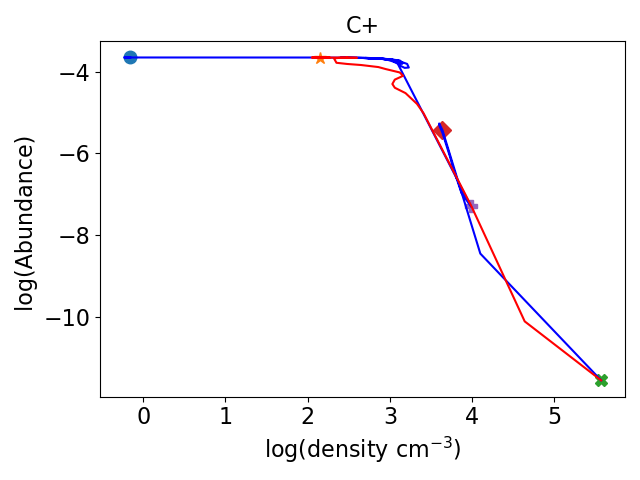}
 \includegraphics[width=0.3\linewidth]{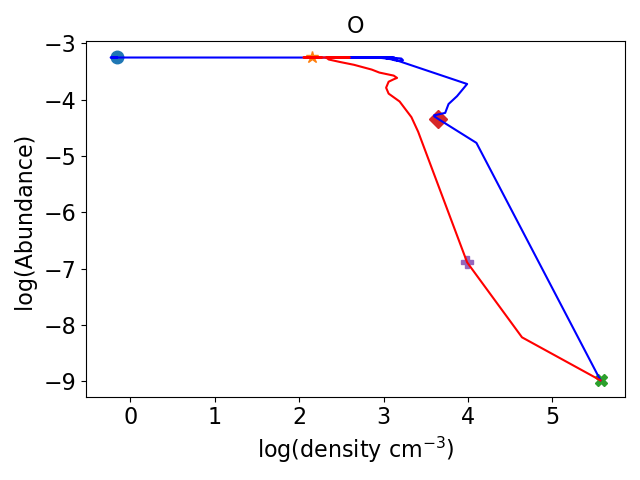}
 \includegraphics[width=0.3\linewidth]{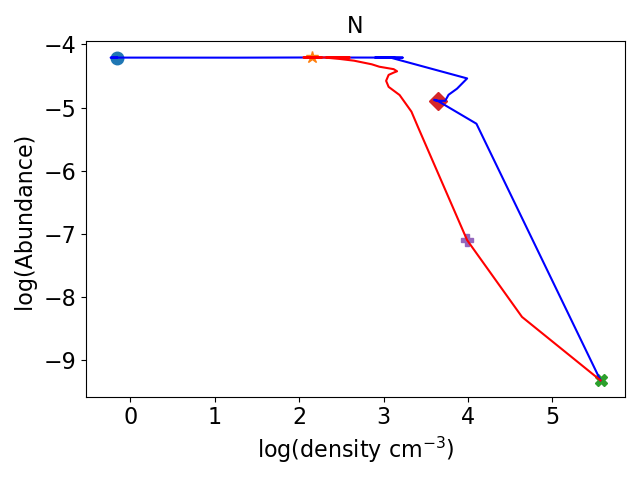}
 \includegraphics[width=0.3\linewidth]{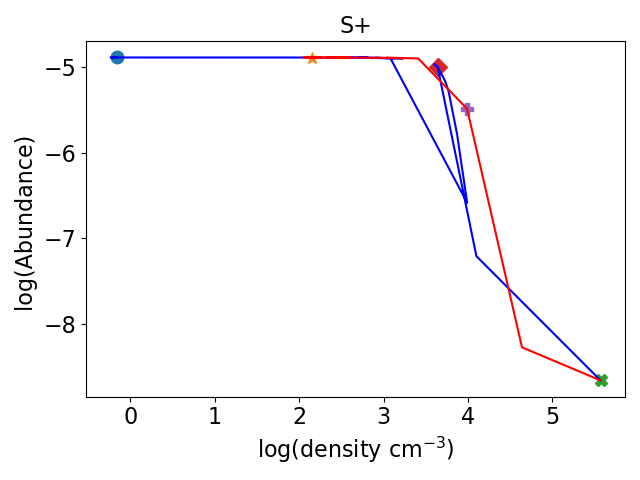}
 \includegraphics[width=0.3\linewidth]{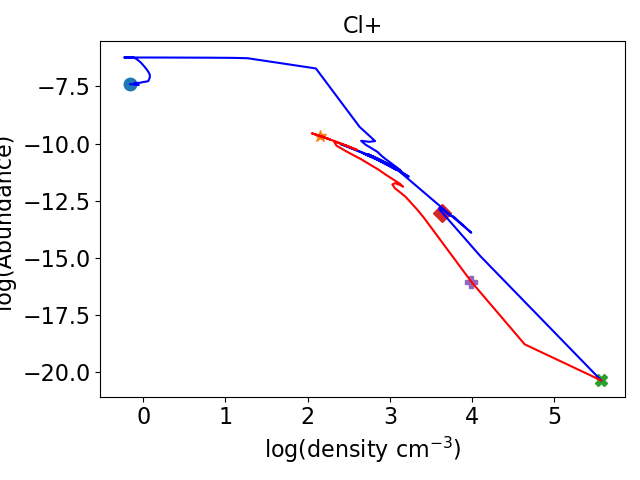}
 \includegraphics[width=0.3\linewidth]{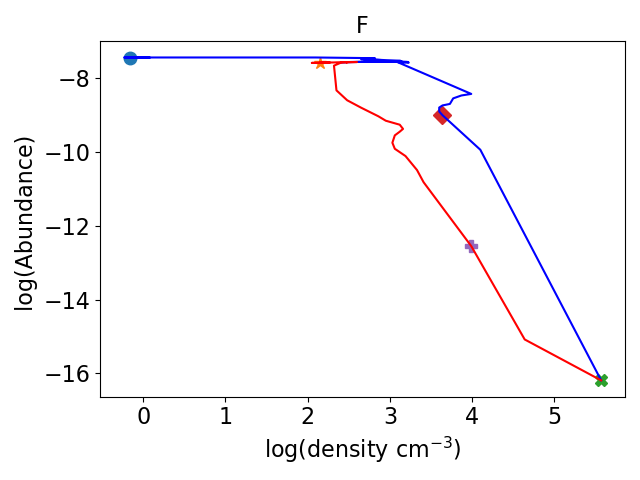}
 \includegraphics[width=0.3\linewidth]{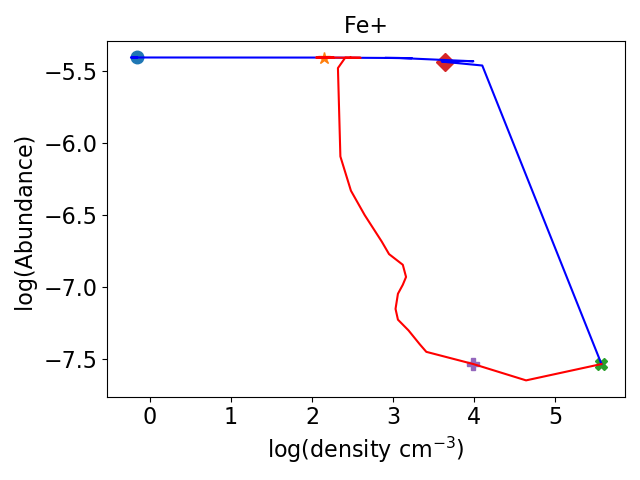}
 \includegraphics[width=0.3\linewidth]{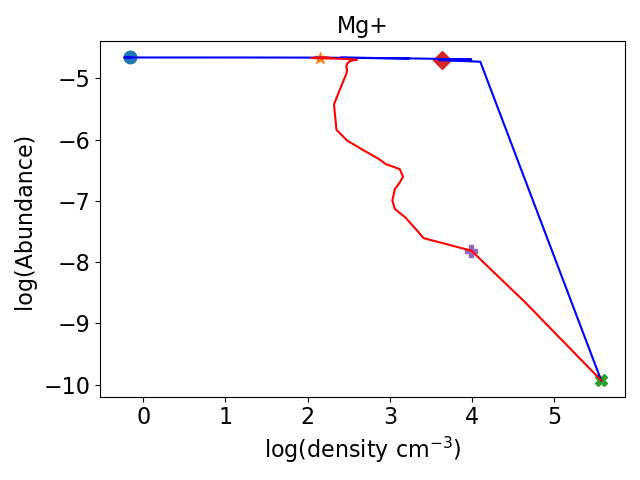}
 \includegraphics[width=0.3\linewidth]{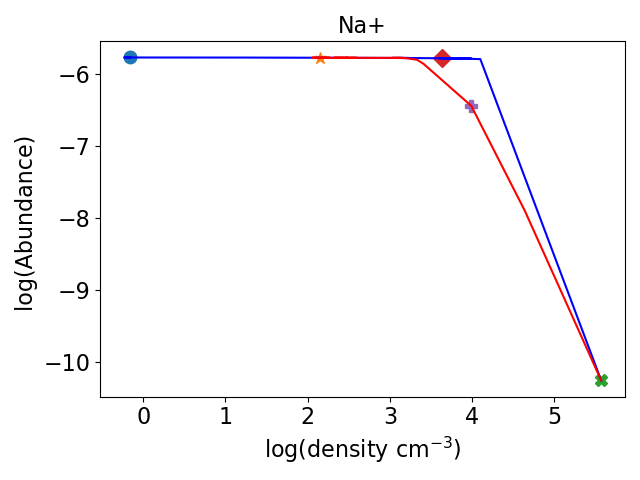}
 \includegraphics[width=0.3\linewidth]{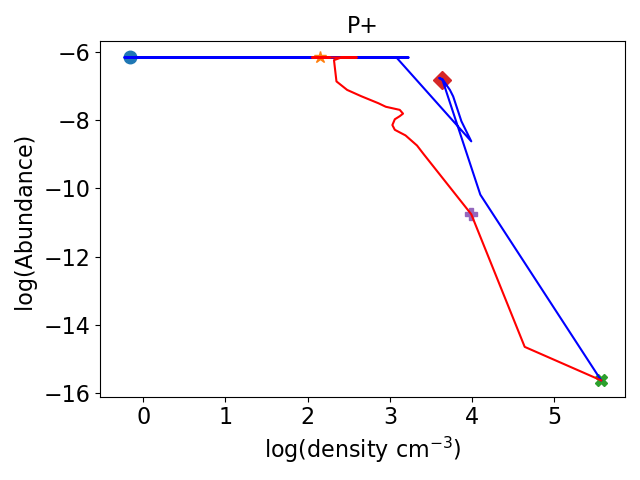}
 \includegraphics[width=0.3\linewidth]{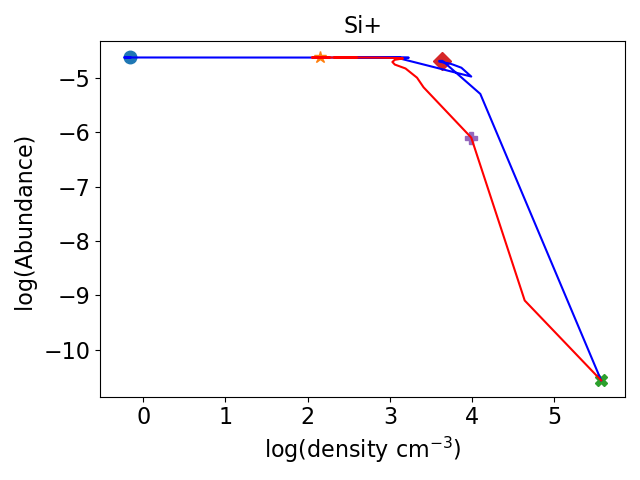}
  \caption{Density as a function of time for trajectory A (upper left panel) and gas-phase atomic abundances (X) as a function of density for the same trajectory for the rest of the figure. Symbols are just markers to help reading the time dependency of the figures. Blue and red parts of the lines represent the phase of increasing and decreasing density. \label{example_trajectoryA}}
\end{figure*}
\begin{figure*}
 \includegraphics[width=0.3\linewidth]{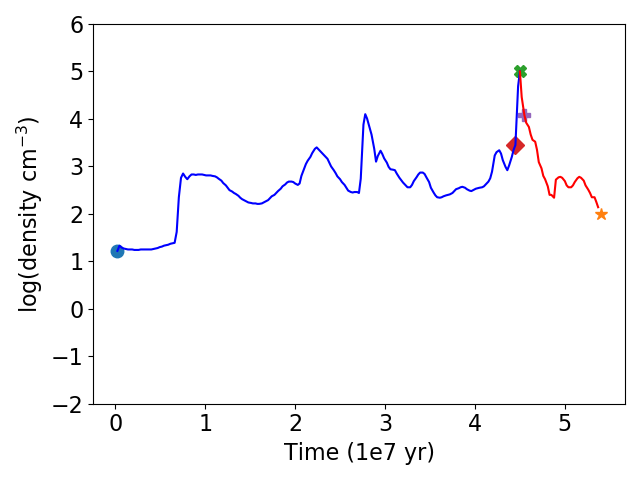}
 \includegraphics[width=0.3\linewidth]{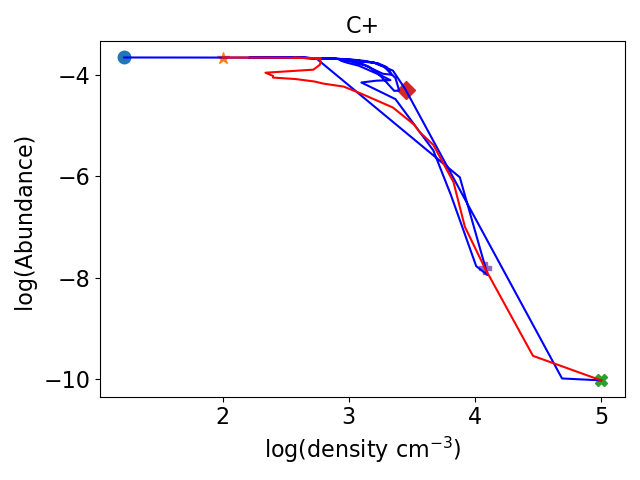}
 \includegraphics[width=0.3\linewidth]{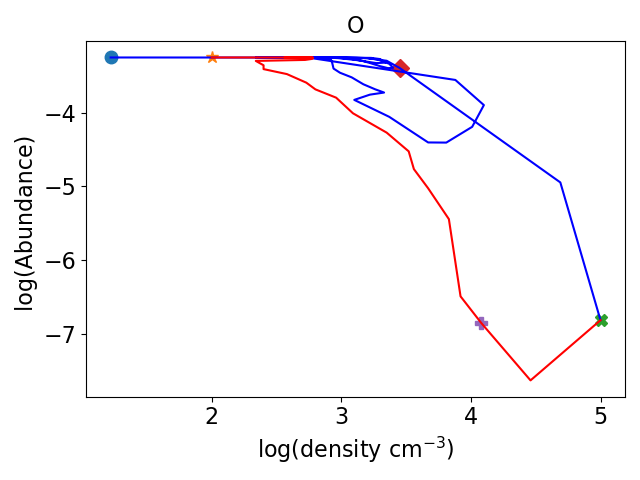}
 \includegraphics[width=0.3\linewidth]{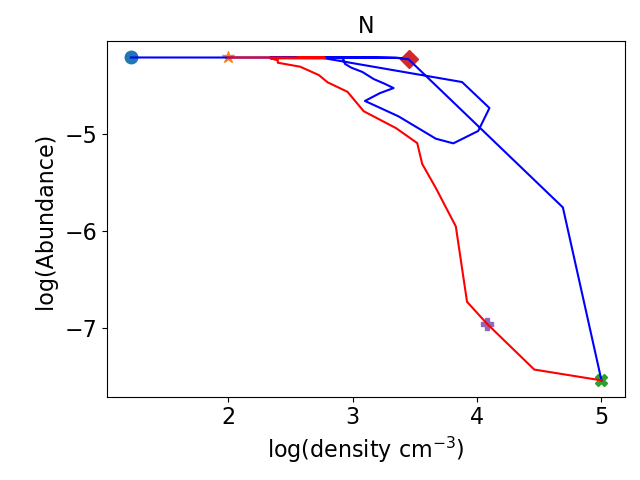}
 \includegraphics[width=0.3\linewidth]{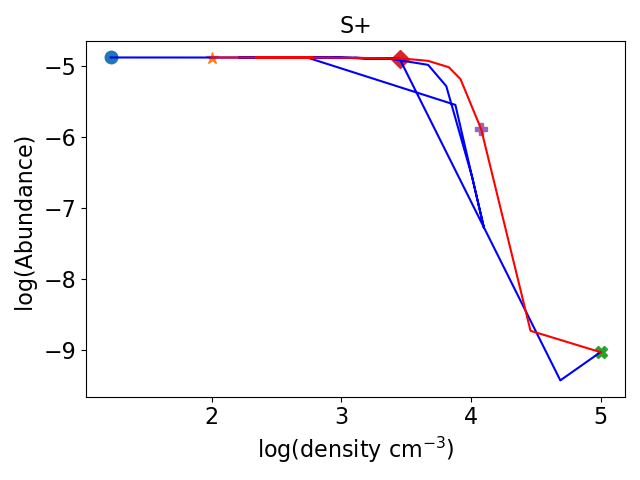}
 \includegraphics[width=0.3\linewidth]{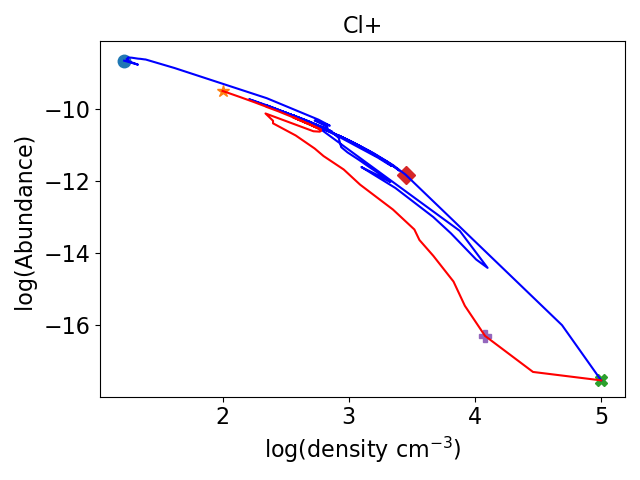}
 \includegraphics[width=0.3\linewidth]{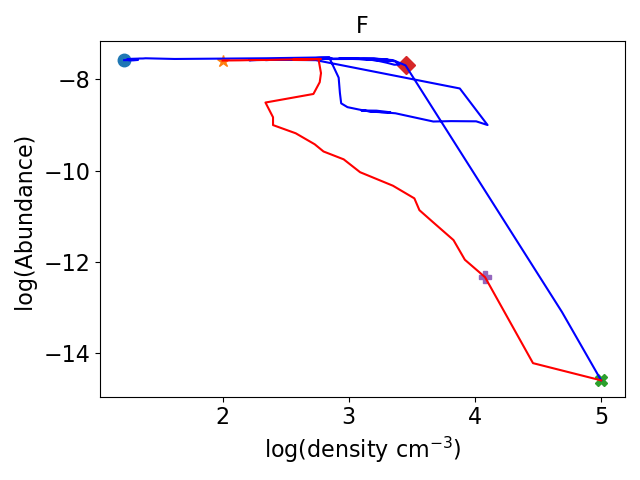}
 \includegraphics[width=0.3\linewidth]{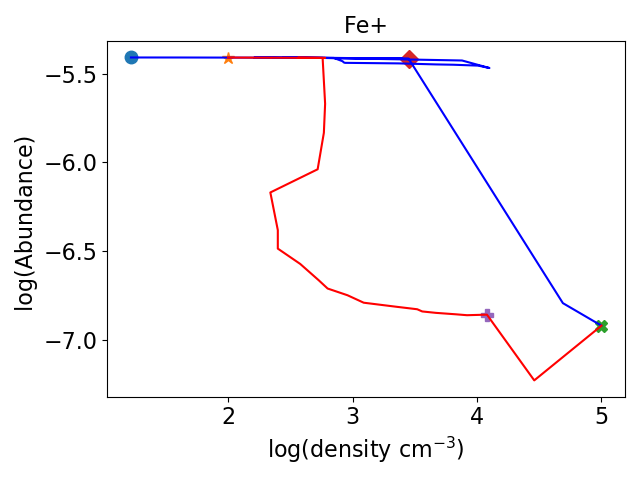}
 \includegraphics[width=0.3\linewidth]{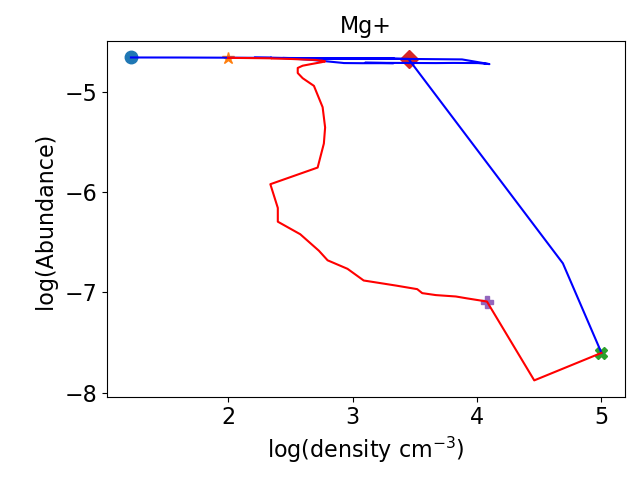}
 \includegraphics[width=0.3\linewidth]{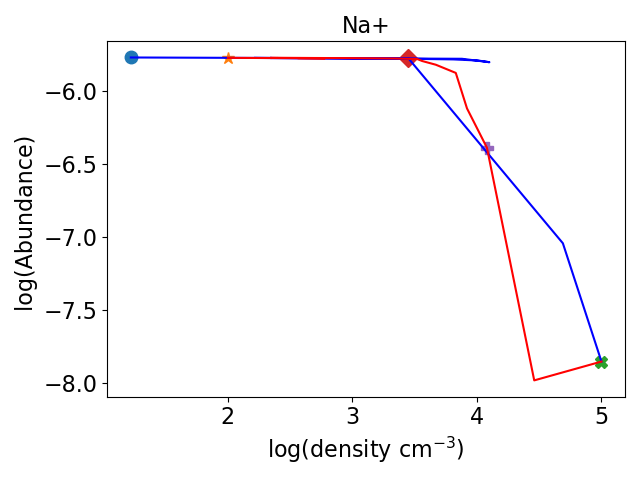}
 \includegraphics[width=0.3\linewidth]{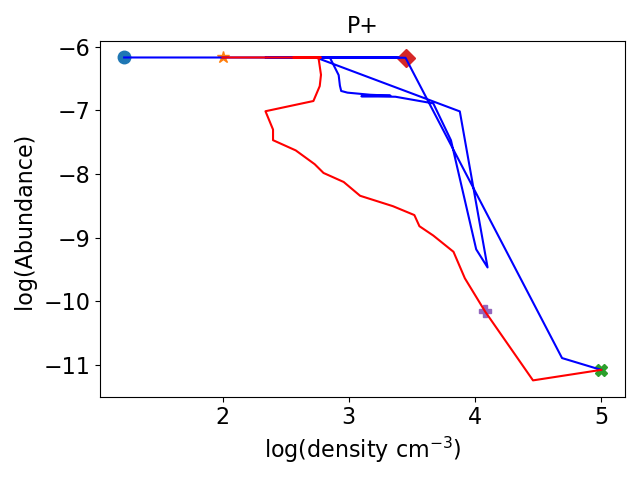}
 \includegraphics[width=0.3\linewidth]{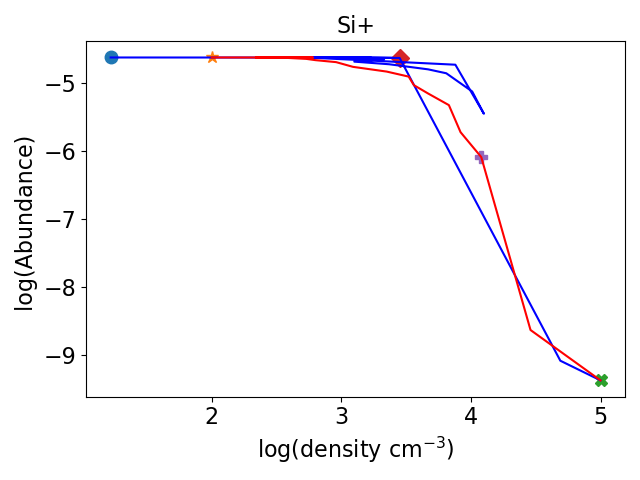}
  \caption{Density as a function of time for trajectory B (upper left panel) and gas-phase atomic abundances (X) as a function of density for the same trajectory for the rest of the figure. Symbols are just markers to help reading the time dependency of the figures. Blue and red parts of the lines represent the phase of increasing and decreasing density.  \label{example_trajectoryB}}
\end{figure*}

\begin{figure}
 \includegraphics[width=1\linewidth]{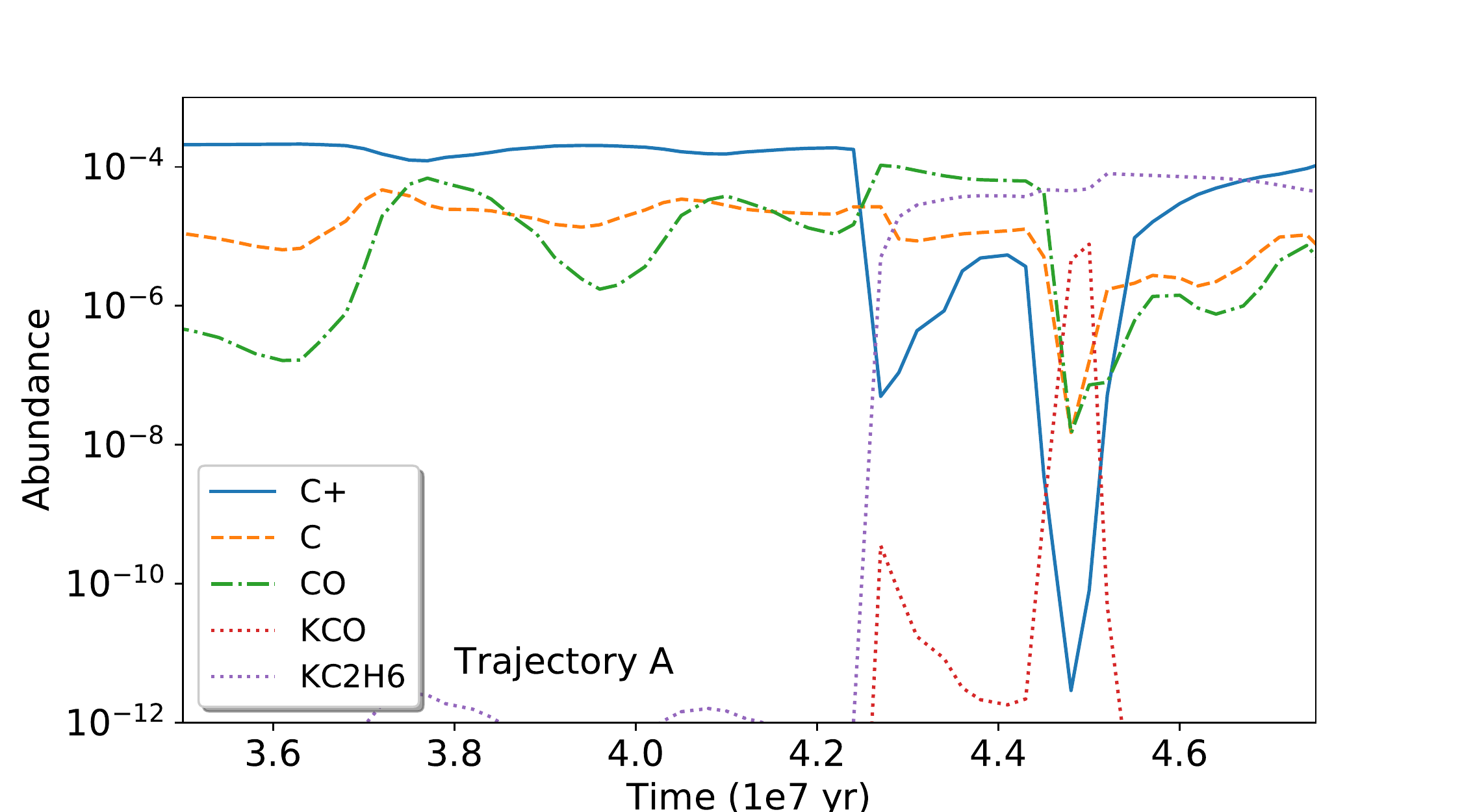}
 \includegraphics[width=1\linewidth]{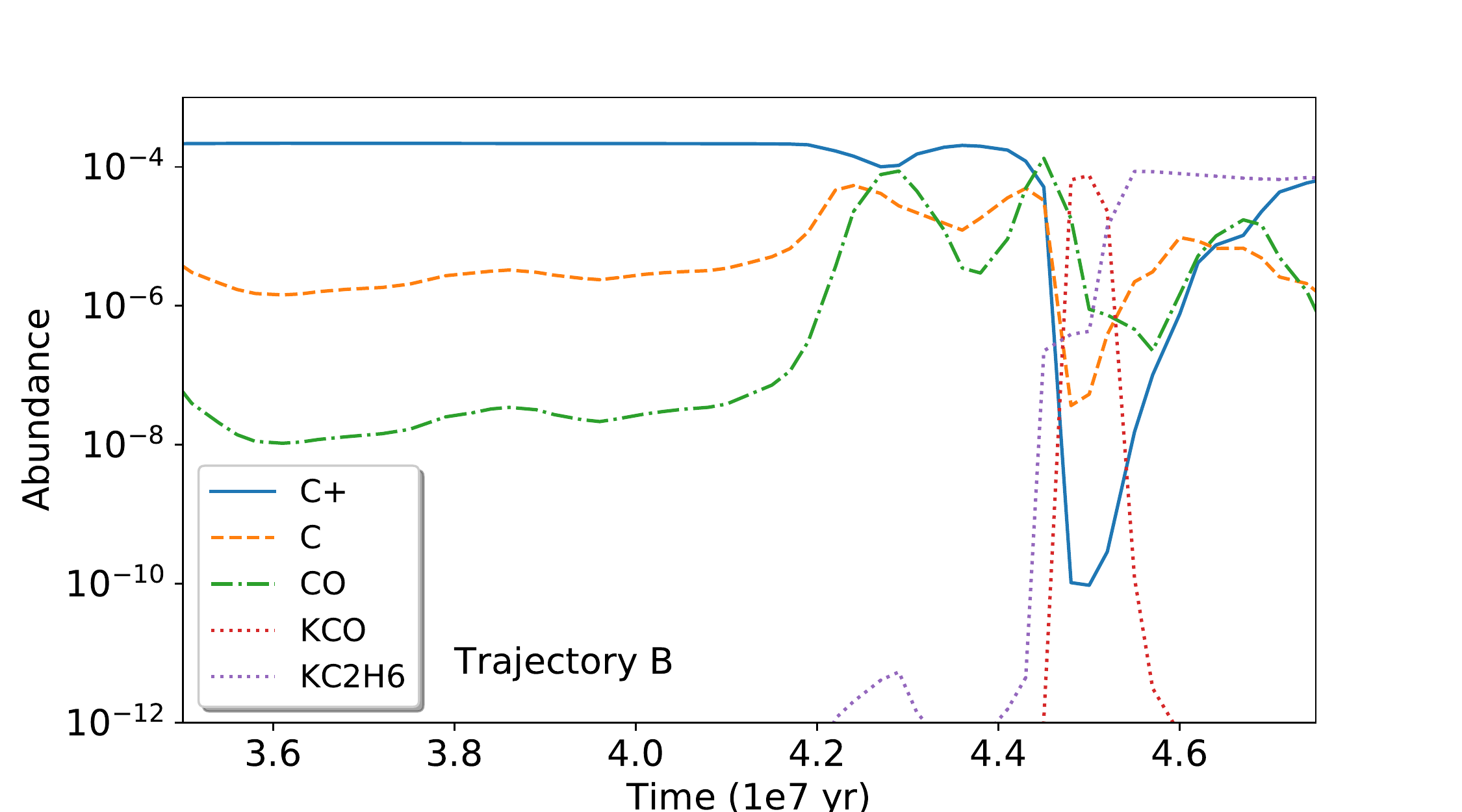}
   \caption{Abundances of the main carbon carriers as a function of time for trajectories A (upper panel) and B (lower panel). K means species in the ices. \label{trajectoriesAB_C}}
\end{figure}

\begin{figure}
 \includegraphics[width=1\linewidth]{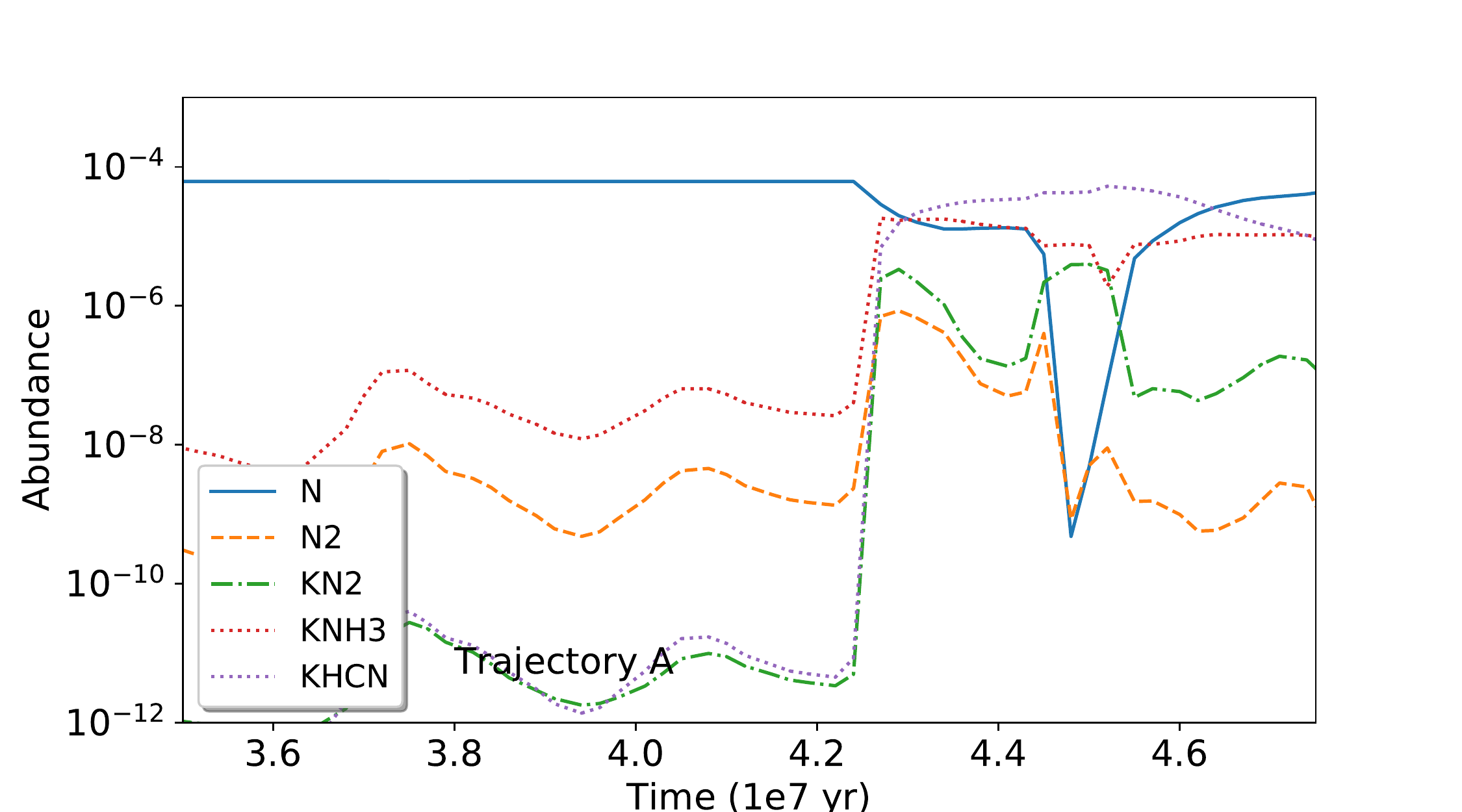}
 \includegraphics[width=1\linewidth]{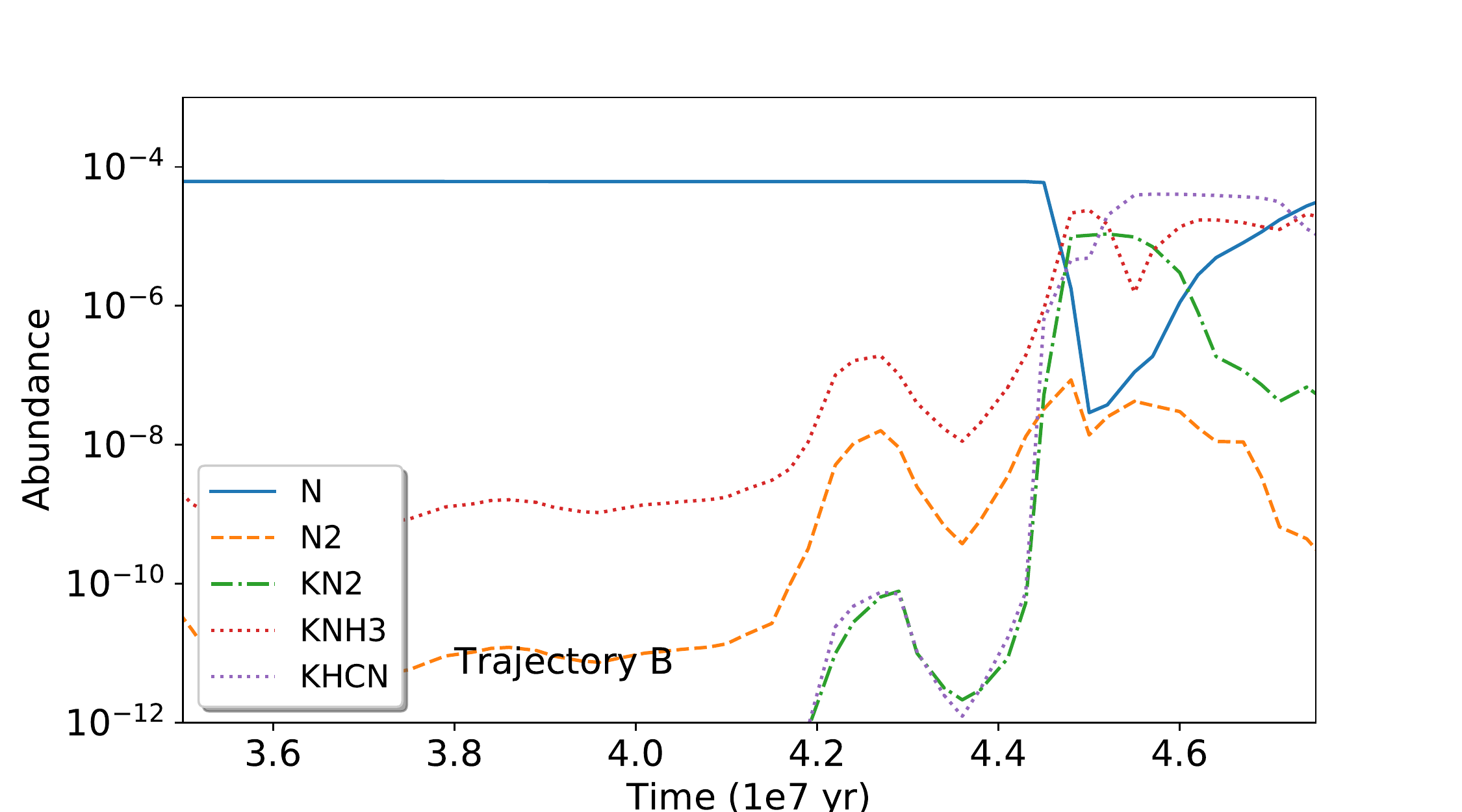}
   \caption{Abundances of the main nitrogen carriers as a function of time for trajectories A (upper panel) and B (lower panel). K means species in the ices. \label{trajectoriesAB_N}}
\end{figure}

\begin{figure}
 \includegraphics[width=1\linewidth]{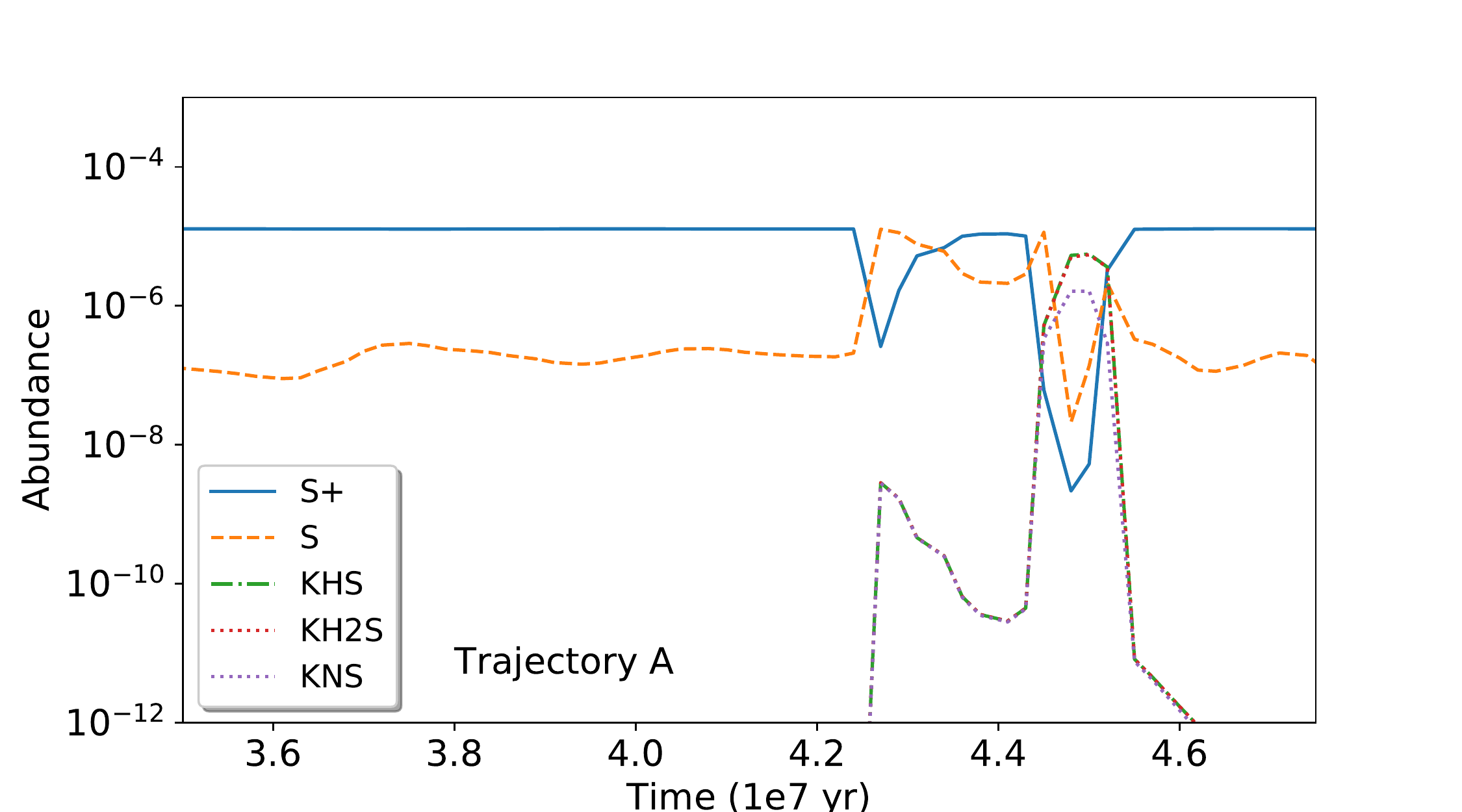}
 \includegraphics[width=1\linewidth]{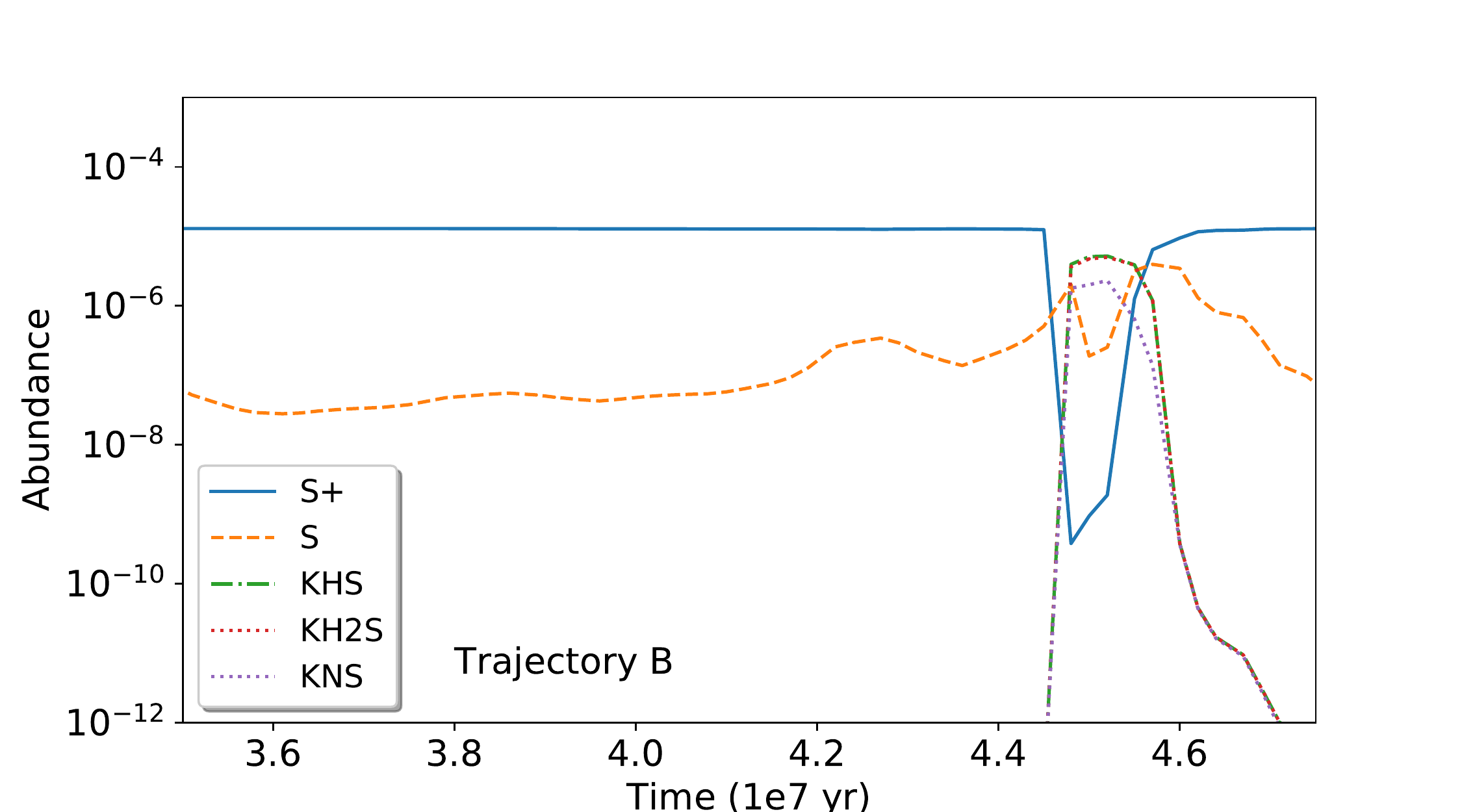}
   \caption{Abundances of the main sulphur carriers as a function of time for trajectories A (upper panel) and B (lower panel). K means species in the ices. \label{trajectoriesAB_S}}
\end{figure}

\begin{figure}
 \includegraphics[width=1\linewidth]{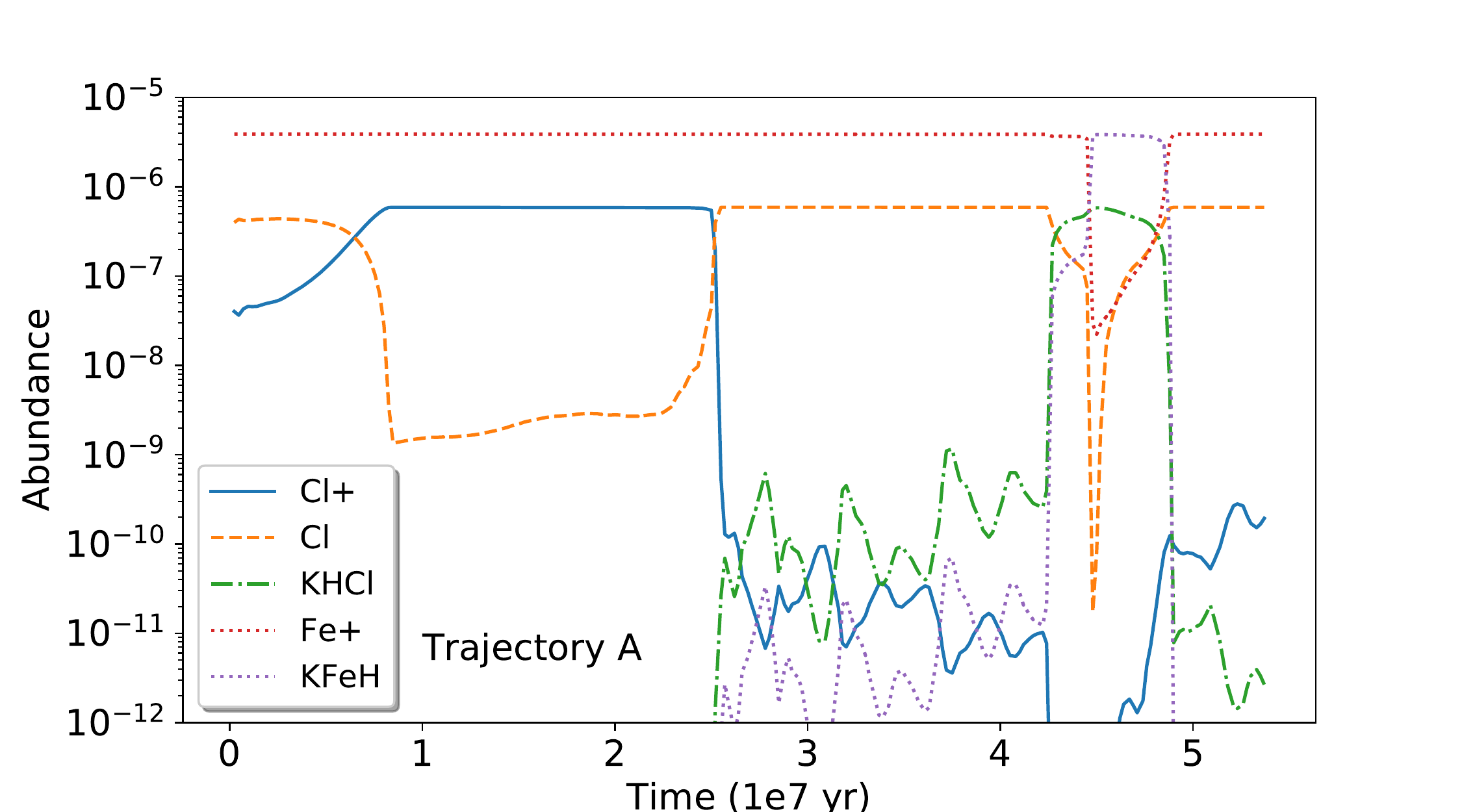}
 \includegraphics[width=1\linewidth]{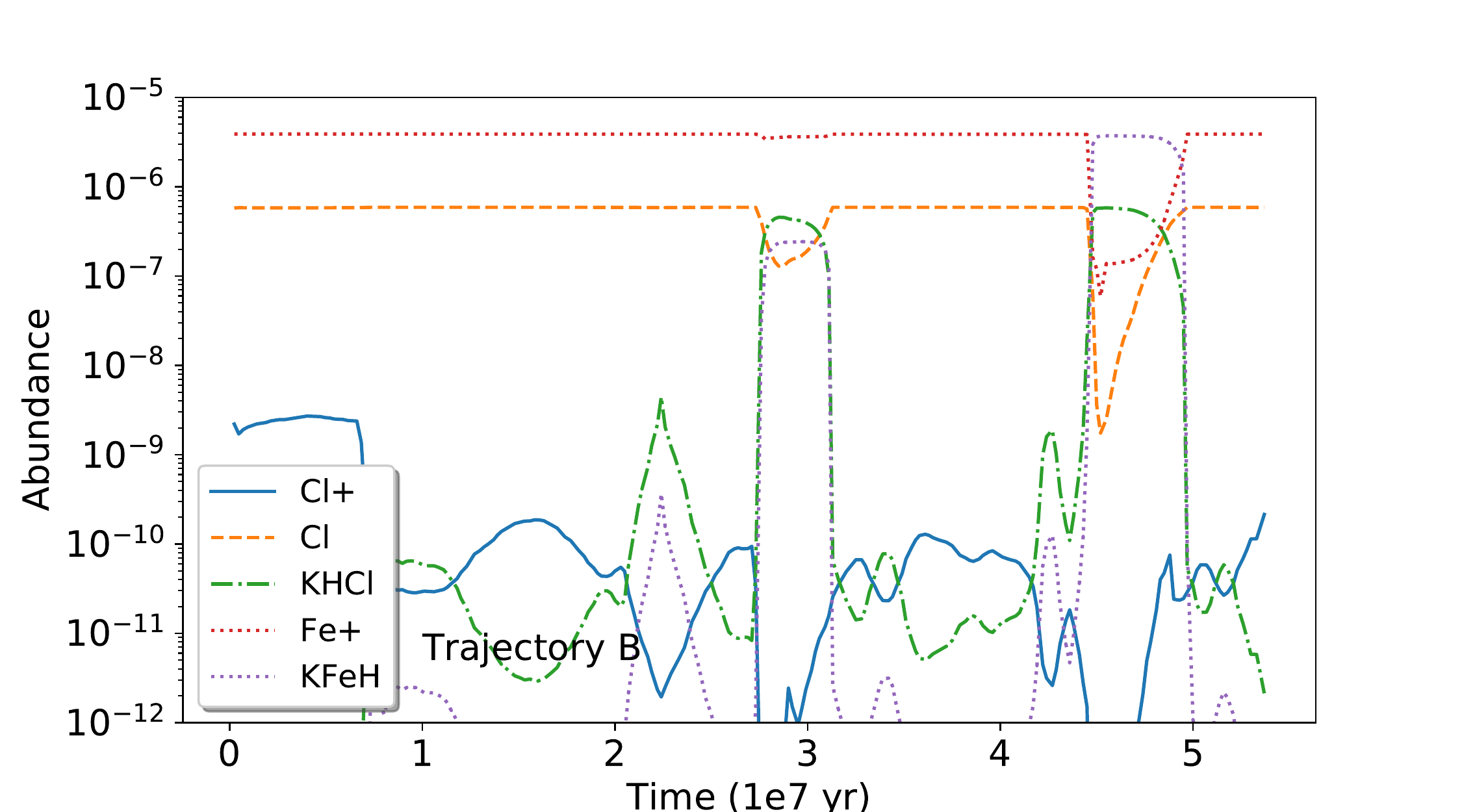}
   \caption{Abundances of the main chlorine and iron carriers as a function of time for trajectories A (upper panel) and B (lower panel). K means species in the ices. Note that the x axis is not the same as for Figs.~\ref{trajectoriesAB_C}, \ref{trajectoriesAB_N}, and \ref{trajectoriesAB_S}.   \label{trajectoriesAB_Cl_Fe}}
\end{figure}


\begin{figure*}
\includegraphics[width=0.43\linewidth]{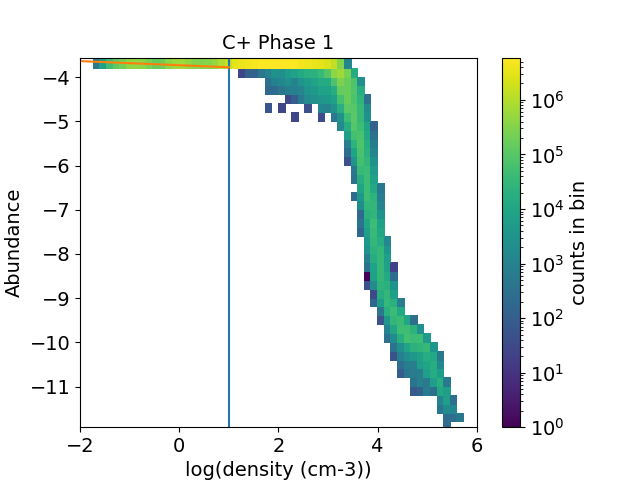}
\includegraphics[width=0.43\linewidth]{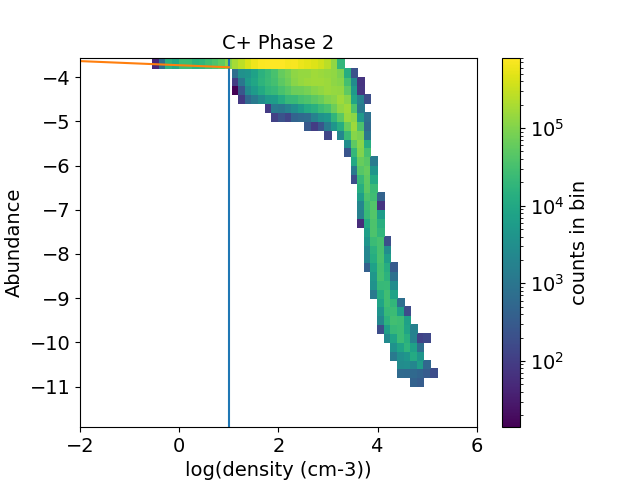}
\includegraphics[width=0.43\linewidth]{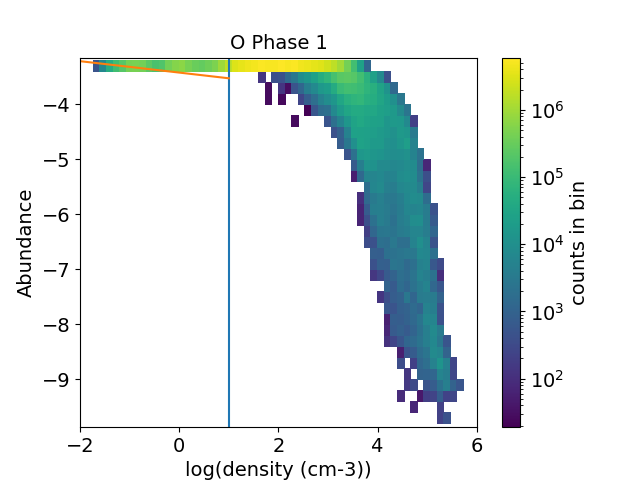}
\includegraphics[width=0.43\linewidth]{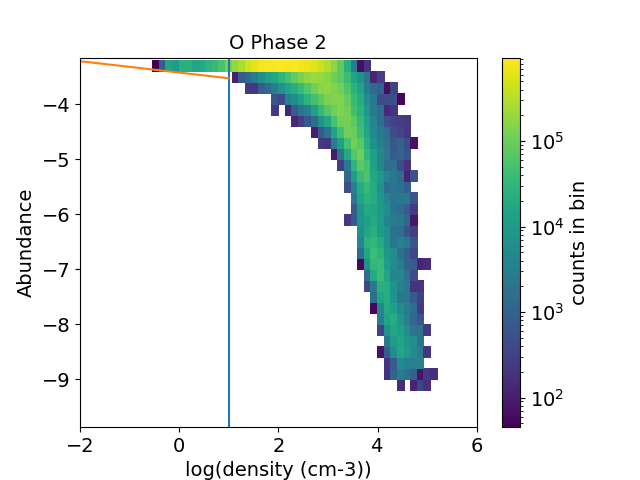}
\includegraphics[width=0.43\linewidth]{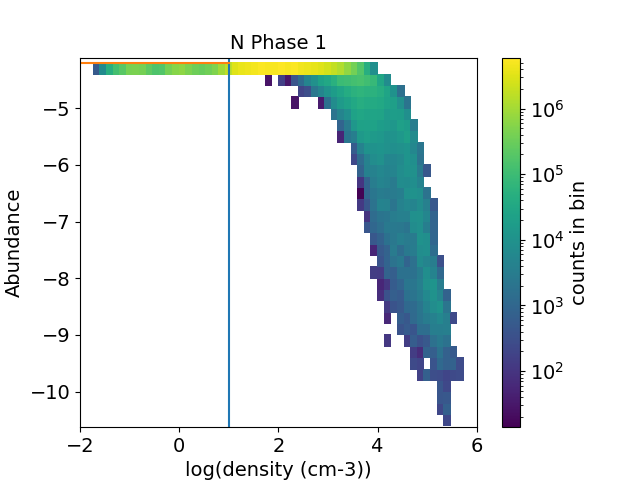}
\includegraphics[width=0.43\linewidth]{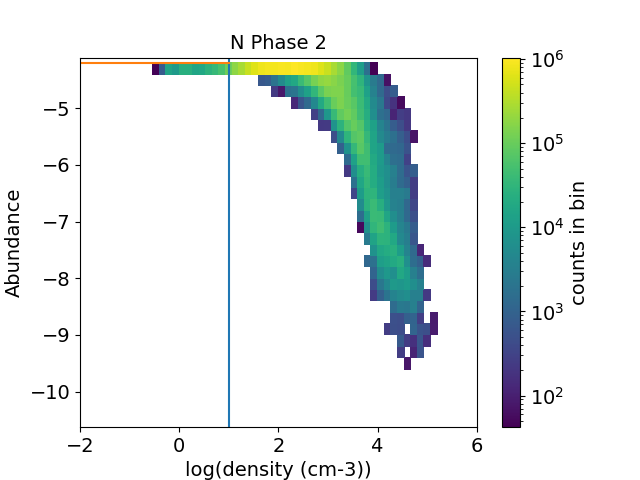}
\includegraphics[width=0.43\linewidth]{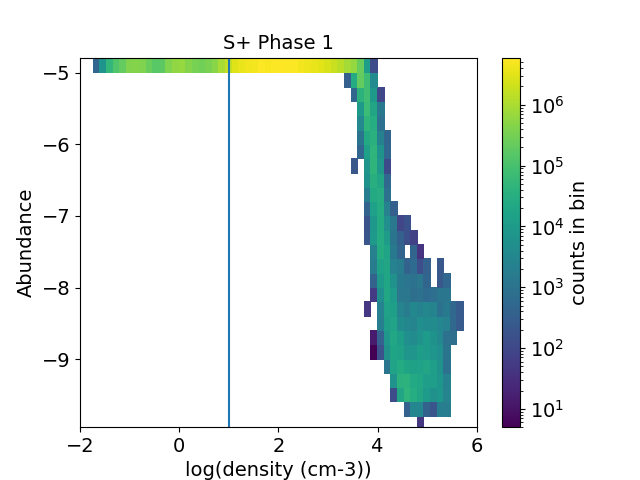}
\includegraphics[width=0.43\linewidth]{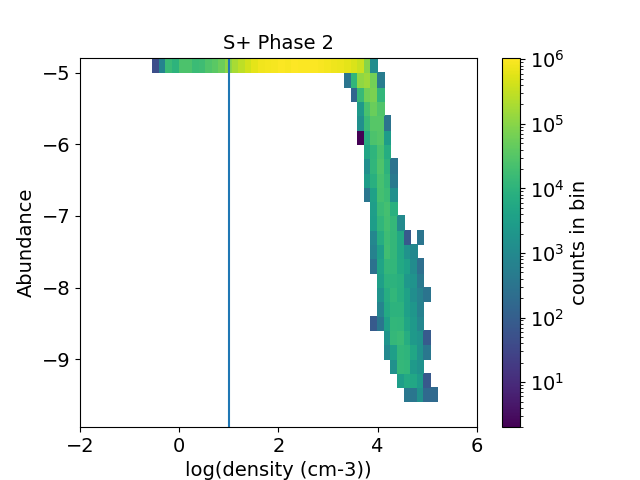}
\caption{Abundances as a function of density for phase 1 (left - (increasing phase of density)) and phase 2 (right - (decreasing phase of density)). The vertical line locates the 10~cm$^{-3}$ density. The red lines show the depletion laws from \citet{2009ApJ...700.1299J}. \label{ab_atoms_1}}
\end{figure*}

\begin{figure*}
\includegraphics[width=0.43\linewidth]{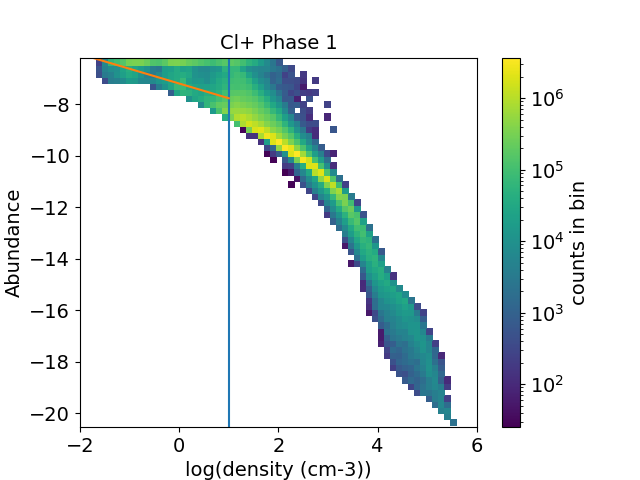}
\includegraphics[width=0.43\linewidth]{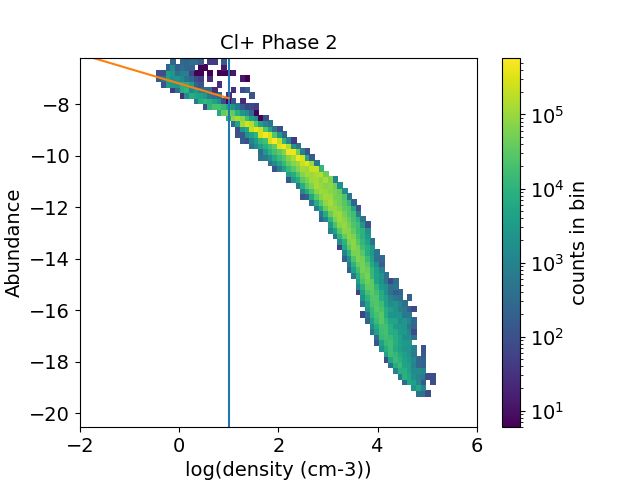}
\includegraphics[width=0.43\linewidth]{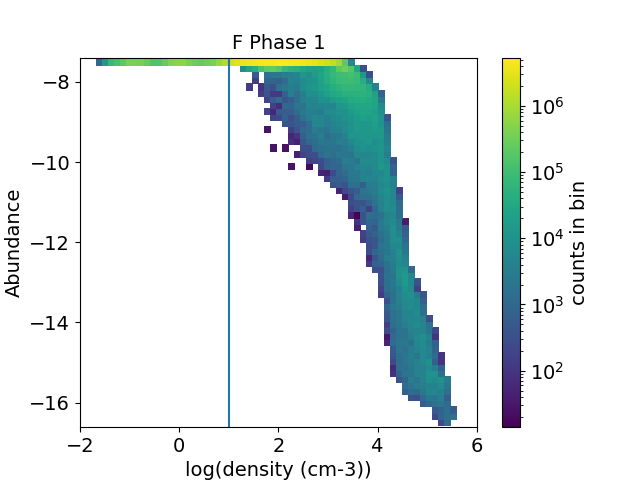}
\includegraphics[width=0.43\linewidth]{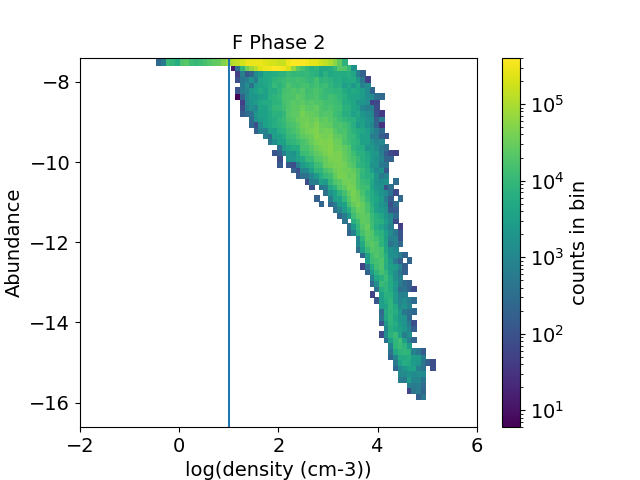}
\includegraphics[width=0.43\linewidth]{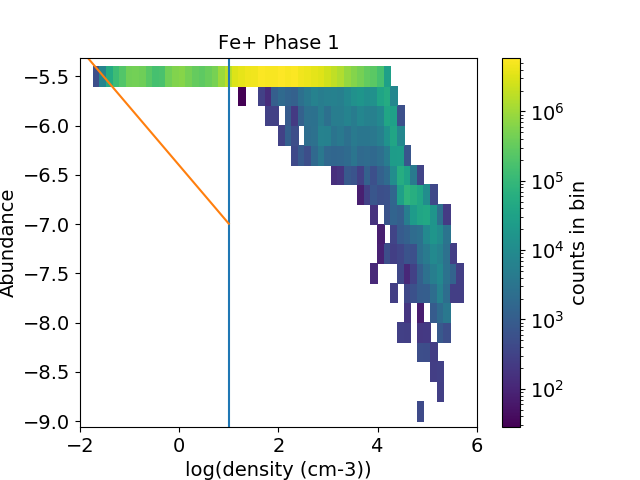}
\includegraphics[width=0.43\linewidth]{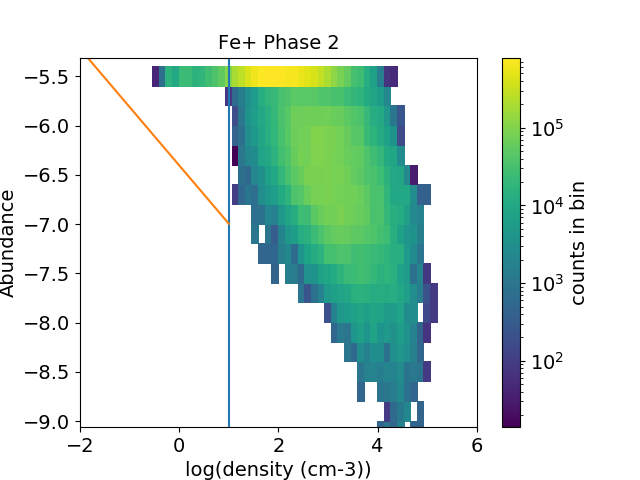}
\includegraphics[width=0.43\linewidth]{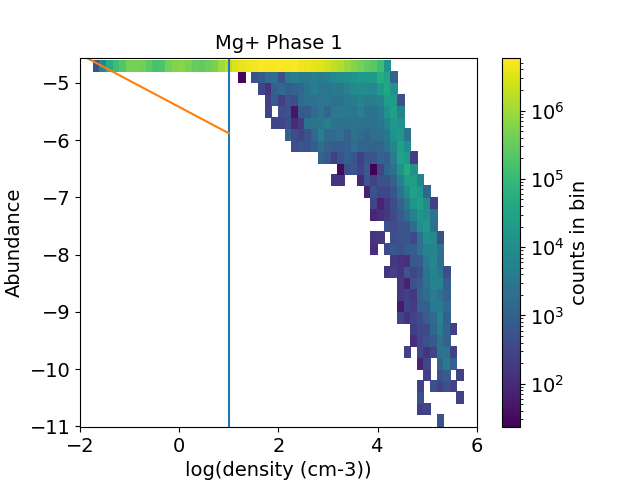}
\includegraphics[width=0.43\linewidth]{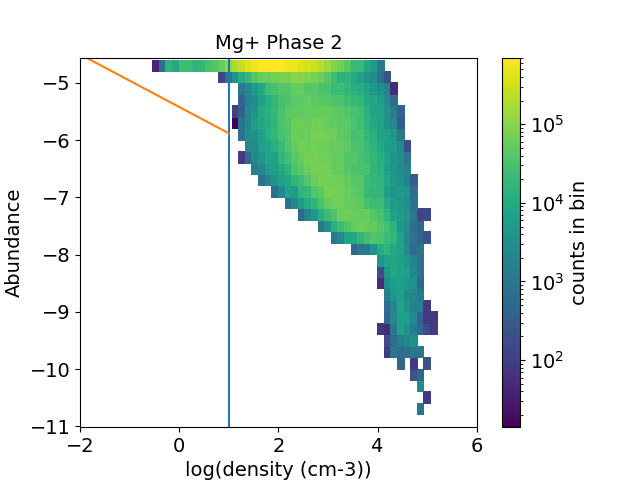}
\caption{Same as figure~\ref{ab_atoms_1}.  \label{ab_atoms_2}}
\end{figure*}

\begin{figure*}
\includegraphics[width=0.43\linewidth]{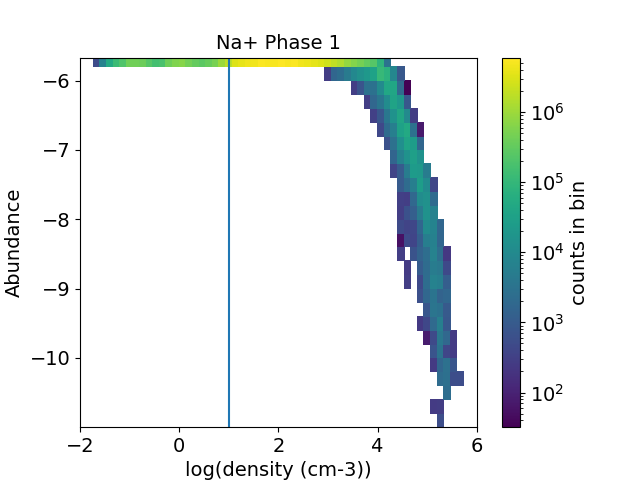}
\includegraphics[width=0.43\linewidth]{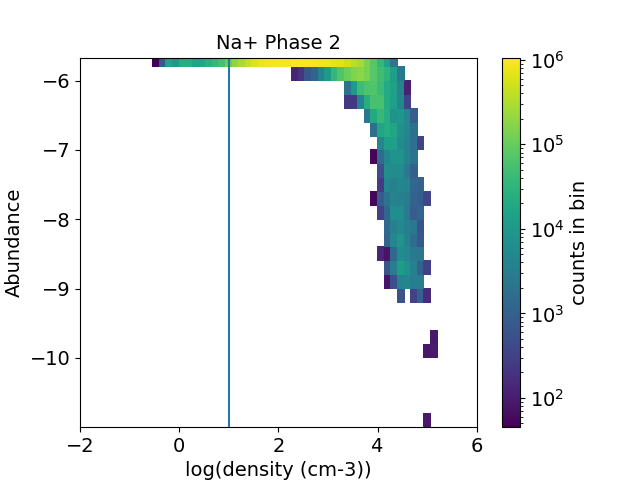}
\includegraphics[width=0.43\linewidth]{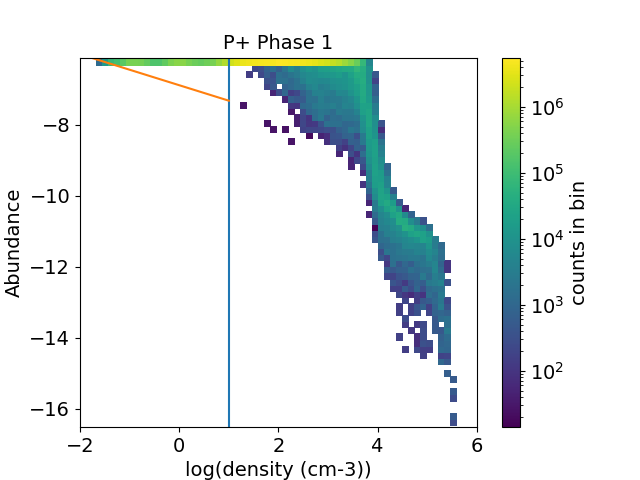}
\includegraphics[width=0.43\linewidth]{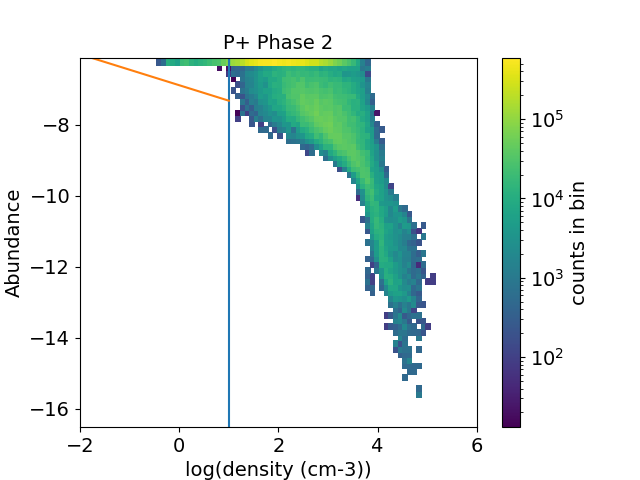}
\includegraphics[width=0.43\linewidth]{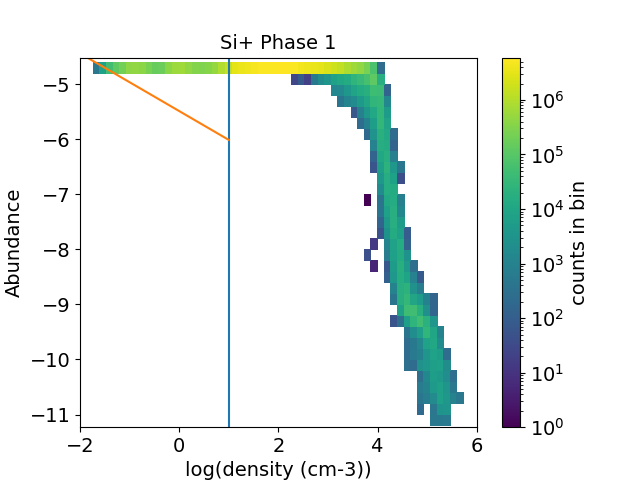}
\includegraphics[width=0.43\linewidth]{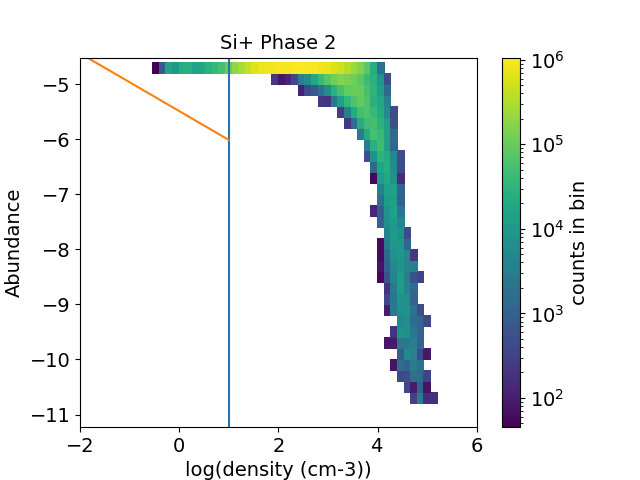}
\caption{Same as figure~\ref{ab_atoms_1}. \label{ab_atoms_3}}
\end{figure*}

In our simulations, we obtain the chemical evolution for thousands of parcel of material, each experiencing different physical conditions, that changes with time. It is not possible to describe the behavior of the species and the chemical processes at play for each of them as we have a large spread of the chemical compositions as was shown in previous publications \citep{2018A&A...611A..96R,2019MNRAS.486.4198W}. We will then first show the examples for two selected trajectories and give more general statistical results considering all of them in a second subsection.

\subsection{Two individual trajectories}

To visualize the results, we first show in Figs.~\ref{example_trajectoryA} and \ref{example_trajectoryB} the density as a function of time for two selected trajectories and the associated atomic gas-phase abundances as a function of density. These two trajectories are the same as in \citet{2019MNRAS.486.4198W}. The dust and gas temperatures as a function of of time for the two trajectories are shown in Appendix~\ref{temperatures}.
On the figures, we use markers to help identifying the initial and final times, as well as the peak density (at $4.45\times 10^7$~yr for both). The time step of physical changes is given by the SPH model ($2.35\times 10^5$~yr). The maximum pic density is then not much resolved representing only a few points. To test the effect of this, we have doubled the number of temporal points by interpolating between two SPH points and run again the model for a few trajectories. The results are not changed significantly.  In these two examples presented here, as in all the other trajectories, the evolution of the density is not linear, resulting in chaotic profiles when looking at the abundance of the atoms as a function of density. The two trajectories do not have the same history of density resulting in a difference in the atomic abundances as a function of density. The starting density of A (0.7 cm$^{-3}$) is smaller than for B (9 cm$^{-3}$), the maximum density is $3.7\times 10^5$~cm$^{-3}$ for A while it is $9.3\times 10^4$~cm$^{-3}$ for B, and the final density is 125 cm$^{-3}$ for A and 56 cm$^{-3}$ for B. 
For both trajectories, except for Cl$^+$, the atomic abundances are rather flat until the density reaches a few $10^3$ or a few $10^4$~cm$^{-3}$ depending on the species. Then the abundances decrease before increasing again when the density decreases. 
 All elements show a minimum at the density peak or just after. In the cases where the minimum abundances do not coincide with the maximum density, the depletion timescale is longer than the evolution of the density. 
 The time scale for the species to come back into the gas-phase depends on the physical conditions, i.e. different for each trajectory. The chaotic behavior of the results makes it difficult to describe generic results concerning the chemistry. 
 
 For elements with complex chemistry, such as carbon, oxygen, nitrogen and sulphur, the elements can be spread over many different species and these species will not be the same for all conditions because it depends on the history \citep{2018A&A...611A..96R}. We show in Fig.~\ref{trajectoriesAB_C} the abundance of the main carriers of carbon as a function of time for both trajectories. In trajectory A, between $2.5\times 10^7$ and $4.2\times 10^7$~yr, C$^+$ dominates, but C and CO are also abundant. During this time, in trajectory B, C and CO are much less abundant than in A, probably because the density is smaller. In A however, the increase of the density at $4.2\times 10^7$~yr is faster than in B so that the C$^+$ abundance drops faster while it remains high for a longer period of time in B. At the same time (when C$^+$ drops), gas-phase CO becomes the major carrier of carbon for approximately $2\times 10^6$~yr in trajectory A. Then at about $4.5\times 10^7$~yr, most of the carbon is locked into C$_2$H$_6$ ice. In trajectory B, at $\sim 4.4\times 10^7$~yr, most of the carbon is first in CO ice for a small amount of time and then in C$_2$H$_6$ ice. In both models, C$^+$ becomes a reservoir again when the density decreases. Oxygen is less complicated as atomic oxygen and water carry the majority of the element. The drop of gas-phase atomic oxygen happens later (after $4.4\times 10^7$~yr) in B while it happens at $4.2\times 10^7$~yr in A. The nitrogen gas-phase atomic abundance drops approximately at the same time as for O. Later, NH$_3$ and HCN ices carry most of the nitrogen but not with the same amont and not at the same time for the two trajectories (see Fig.~\ref{trajectoriesAB_N}). For sulphur (Fig.~\ref{trajectoriesAB_S}), we also see a decrease of gas-phase S$^+$ at the same time as O and N but S$^+$ and S alternatively share most of the sulphur between $4.2\times 10^7$ and $4.4\times 10^7$ yr before forming HS and H$_2$S ices for trajectory A. This alternance of S$^+$ and S are not seen for trajectory B. 

For Fe, Mg, Na, and Si, which are initially ionic, the decrease in abundance is first due to the electronic recombination with electrons (this is in fact also the case for C and S). The neutral atoms are then depleted on interstellar grains during collision, and hydrogenated. As such, FeH, MgH$_2$, NaH, and SiH$_4$ ices are the main carriers of these elements at high density. The case of iron is shown in Fig.~\ref{trajectoriesAB_Cl_Fe}. F and P$^+$ react mostly with H$_2$ to form HF and PH$_2^+$. HF then depletes on the grains while PH$_2^+$ reacts with electrons to produce P (and PH but PH gives P through the reactions PH + C$^+$ $\rightarrow$ PH$^+$ + C and PH$^+$ + e$-$ $\rightarrow$ P + H). Atomic neutral phosphorus then depletes on the grains and is hydrogenated into PH$_3$. HF and PH$_3$ ices are the main carriers of fluorine and phosphorus at high density.
 Chlorine appears as an exception here because the Cl$^+$ gas-phase abundance drops at much smaller density (see Fig. \ref{trajectoriesAB_Cl_Fe}) as compared to the other elements. This fast Cl$^+$ $\rightarrow$ Cl conversion is due to the combined effect of the electronic recombination of Cl$^+$ (Cl$^+$ + e$^-$ $\rightarrow$ Cl + h$\nu$) and the Cl$^+$ + H$_2$ $\rightarrow$ HCl$^+$ + H reaction (followed by HCl$^+$ + H$_2$ $\rightarrow$ H$_2$Cl$^+$ + H and H$_2$Cl$^+$ + e$^-$ $\rightarrow$ Cl + 2H). We checked this hypothesis by decreasing each and then both rate coefficients. The rate coefficient of the electronic recombination of Cl$^+$, in our network, is $1.13\times 10^{-10} (T/300)^{-0.7}$~cm$^3$~s$^{-1}$. This rate coefficient is more than one order of magnitude larger than the typical electronic recombination included in astrochemical databases. This rate coefficient is indicated in the KIDA database\footnote{http://kida.obs.u-bordeaux1.fr/} \citep{2012ApJS..199...21W} and comes from the 1991 version of UMIST\footnote{http://udfa.ajmarkwick.net/} \citep{1991A&AS...87..585M}. In the last UMIST version (2012) \citep{2013A&A...550A..36M}, it has been modified towards smaller values but without any reference. More laboratory work is required on this reaction. Decreasing the rate coefficient however does not significantly change the Cl$^+$ abundance as long as the Cl$^+$ + H$_2$ $\rightarrow$ HCl$^+$ + H rate coefficient stays high \citep[$10^{-9}$~cm$^{-3}$~s$^{-1}$,][]{2009ApJ...706.1594N}. The two trajectories show different results for Cl. As the density is initially smaller in A, the gas-phase abundance of Cl$^+$ is high for about $2.5\times 10^7$~yr while it is always low in B. Around $3\times 10^7$~yr, trajectory B experiences a bump in density high enough to produce large amounts of HCl ice but that is then dissociated and brought back into the gas-phase as neutral Cl before the maximum density peak is achieved. At high density, HCl ices are the main carrier of chlorine.\\

During the second phase, after the density peak, for the heavy elements with a simple chemistry (Mg, F, Fe, P, Si, Na, and Cl), the reservoirs stored on the grains are first dissociated on the surfaces as the density decreases. The atoms are then evaporated and ionized (except for fluorine). Photodesorption plays here a minor role. If the efficiency of the process scales with the visual extinction, the yield is small ($10^{-4}$) and the same for all species \citep{2016MNRAS.459.3756R}. Note that the fluorine atomic gas-phase abundance F in the two examples shown in Figs.~\ref{example_trajectoryA} and \ref{example_trajectoryB} only represents 67\% of the elemental abundance at the end of the simulation. The remaining fluorine is in the form of HF in the gas because of a rapid F + H$_2$ $\rightarrow$ H + HF reaction. For smaller densities (as shown in Fig.~\ref{ab_atoms_2}), HF eventually dissociates and F becomes the main fluorine carrier again.

\subsection{All trajectories}\label{alltrajectories}

As time is a model dependent parameter, we plot the results as a function of density in Figs.~\ref{ab_atoms_1} to \ref{ab_atoms_3} for all trajectories. The abundances during the first phase (increasing density) are shown in the left and on the right for the second phase (decreasing density). Considering all trajectories, we find similar behavior as previously shown. For most elements, the decrease in the atomic abundances starts when the density is larger than $\sim 10$~cm$^{-3}$. Chlorine is an exception as its abundance decreases for some trajectories for densities below 0.1 cm$^{-3}$. 
Some of the trajectories do not show any elemental depletion at high density, except for Cl$^+$ which is always depleted. This is because, within the full range of histories studied here, there are always trajectories presenting fast increase of the density, which induces a delay in the molecular depletion. Chlorine again does not present this behavior because of its fast conversion to its neutral form in the gas-phase.
For all elements, the abundances in phase 2 (decreasing phase of density) are smaller at low density than in phase 1 (increasing phase of density), meaning that the depletion experienced in the dense phase impacts the inventory of available gas phase elements in phase 2. This effect is stronger for C$^+$, Fe$^+$, Mg$^+$, P$^+$, Cl$^+$ and F. Later on, the gas abundances of Mg$^+$, P$^+$, F, Si$^+$, Fe$^+$, and Na$^+$ are equal to the initial abundances used in the model when the density is smaller than 10~cm$^{-3}$. \\
On Figs.~\ref{ab_atoms_1} to \ref{ab_atoms_3}, we have superimposed the elemental depletion laws derived by \citet{2009ApJ...700.1299J} from atomic line observations in the diffuse interstellar medium (up to 10~cm$^{-3}$. For carbon, oxygen and nitrogen, the observed elemental abundance is small and not in contradiction with our predictions. For iron, magnesium, phosphorus, and silicon, the observed depletion is much stronger than predicted by the model. For chlorine, we do reproduce the observed depletion.

\section{Discussion and Conclusion}

In this paper, we studied the depletion of the elements during the transition between the diffuse and dense interstellar medium by coupling a full gas-grain chemical model to results obtained from large scale hydrodynamics simulations of the interstellar medium at galactic scales. With this approach we could follow the evolution of interstellar matter over several millions of years and study the effect of cycling between low density and high density phases on the depletion of elements such as C, O, N, S, Si, Fe, Na, Mg, P, Cl and F. 
For all these elements, we find that the strength of their depletion is set by the balance between accretion/reactions at the surface and their ability to resist photoprocessing. Our main result is that all these elements, but Cl, recover their undepleted values when $\rm n_H < 10$~cm$^{-3}$ and our model thus fails at reproducing depletion patterns derived from observations of low density gas  \citep{2009ApJ...700.1299J}. 

Chlorine is an exception and the depletion pattern of Cl$^+$ at $\rm n_H < 10$~cm$^{-3}$ agrees well with the one derived by \citet{2009ApJ...700.1299J}. Because of the fast electronic recombination of Cl$^+$ and its efficient reaction with H$_2$, Cl$^+$ disappears efficiently even at very low density. The fact that we are able to reproduce the chlorine observed depletion at very low density, may indicate that the electronic recombination of other elements might be too low. \citet{2009ApJ...694..286B} have tested the impact of new more accurate estimates of the electronic recombination of He$^+$, C$^+$, O$^+$, Na$^+$, and Mg$^+$ for a variety of physical conditions. According to their calculations, the largest difference between the new rate coefficients and the ones typically listed in current astrochemical databases was for for Mg$^+$ electronic recombination, which is 60\% smaller at 10~K in the new estimates. Such differences may not impact our calculations. However, it could be worth deriving parametrized rate coefficients from the original calculations to include them in astrochemical models. The authors also stated that there exists no reliable data for Si$^+$, P$^+$, S$^+$, Cl$^+$, and Fe$^+$. 

Possible explanations for discrepancies observed at low densities may also come from unaccounted processes in the current chemical model. 
A first possibility comes from the single grain size approximation we make. For instance, a MRN size distribution extended down to $\rm a_{min} \sim10$~\AA decreases the collision timescale between atoms and grains by a factor of $\sim$6 (i.e. t$\sim 10^7$~yrs at $\rm n_H \sim 10$~cm$^{-3}$). If we further take into account that a significant fraction of these small grains (which dominate the surface area) would be negatively charged, Coulomb focusing reduces this timescale to few $10^5$~yrs making possible the in-situ depletion of some of the atomic cations \citep{1999ApJ...517..292W}. The non-thermal desorption mechanism (though cosmic-ray heating for instance) would however reduce strongly the depletion effect of the small grains \citep{2006MNRAS.367.1757C}.

However, additional considerations such as the strength of the adsorption at the surface should also be explored. Even though enhanced collision rates could facilitate the depletion process, a fraction of these elements must remain bound to the surface at very low densities (this is especially true for elements such as Mg, Fe, P and Si, for which significant depletion has been inferred at low densities). In the present study we assumed physisorption, for which typical binding energy vary between 10 and few 100 meV. Such low binding energies lead relatively fast thermal desorption timescale under diffuse cloud conditions. Indeed, we find that the combination of photoprocessing and thermal evaporation is efficient for clearing-out the grains surface from any adsorbed atoms and molecules on timescales of few $10^4$~yrs at $\rm n_H \sim 10$~cm$^{-3}$. The inclusion of chemisorption sites, characterized by increased binding energies ($> 0.1$~eV), may be able to keep a significant fraction of these elements bound to the surface under diffuse cloud conditions and should thus be included in future studies. However, if chemisorption has been studied for hydrogen atoms, to the best of our knowledge, there exist little information on this process for other elements. In particular, the strength of chemisorption under high irradiation conditions remain to be studied in details.

On the other hand, if we assume that the diffuse ISM observed depletion is due to depletion in dense regions that have been brought back into diffuse conditions, another possibility comes from the production of refractory material resulting from irradiation of the ice mantles during the transition between dense molecular clouds to diffuse clouds. Laboratory experiments on the irradiation of interstellar ice analogues have demonstrated an efficient formation of refractory compounds \citep[e.g. ][]{1995ApJ...454..327B,2011AdSpR..48.1126N,2015BSRSL..84...21A} and in particular material very similar to insoluble organic material (IOM) found in meteorites and IDPs \citep{2011AdSpR..48.1126N}. These residues should be relatively refractory to processes such as thermal evaporation and photodesorption (even though they may be altered by UV irradiation) under diffuse clouds conditions and could thus represent possible candidate for interstellar depletion carriers. Chemical networks that include such complexity remain to be developed.

\section*{Acknowledgments}

VW's research was funded by an ERC Starting Grant (3DICE, grant agreement 336474). The authors acknowledge the CNRS program "Physique et Chimie du Milieu Interstellaire" (PCMI) co-funded by the Centre National d'Etudes Spatiales (CNES). IAB acknowledges funding from the European Research Council for the FP7 ERC advanced grant project ECOGAL. This work used the DiRAC Complexity system, operated by the University of Leicester IT Services, which forms part of the STFC DiRAC HPC Facility (www.dirac.ac.uk). This equipment is funded by BIS National E-Infrastructure capital grant ST/K000373/1 and STFC DiRAC Operations grant ST/K0003259/1. DiRAC is part of the National E-Infrastructure. MR\~{o}s research was supported by an appointment to the NASA Postdoctoral Program at the NASA Ames Research Center, administered by Universities Space Research Association under contract with NASA.

\section*{Data Availability}
The physical and chemical simulations and the nautilus gas-grain model are available upon request.





\bibliographystyle{mnras}
\bibliography{biblio}



\appendix

\section{Dust and gas-temperatures for trajectories A and B}\label{temperatures}

\begin{figure}
 \includegraphics[width=1\linewidth]{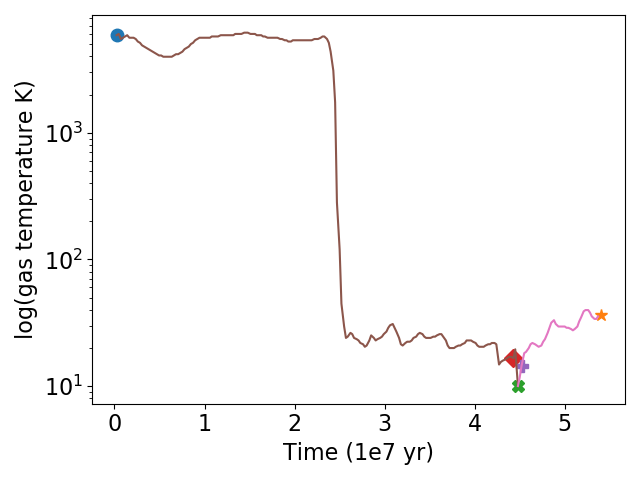}
 \includegraphics[width=1\linewidth]{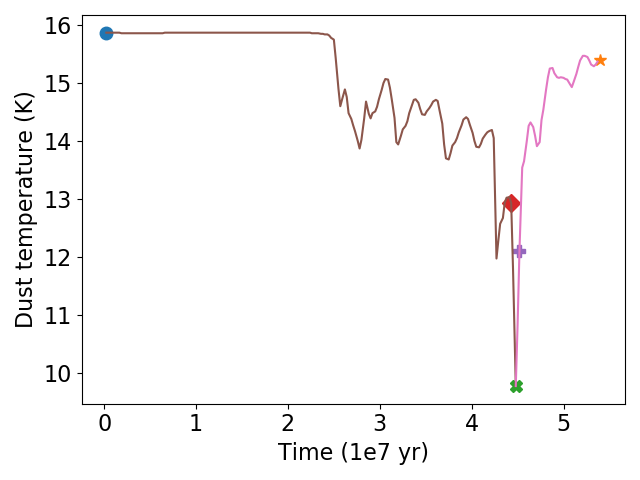}
   \caption{Gas and dust temperature as a function of time for trajectory A.}
\end{figure}
\begin{figure}
 \includegraphics[width=1\linewidth]{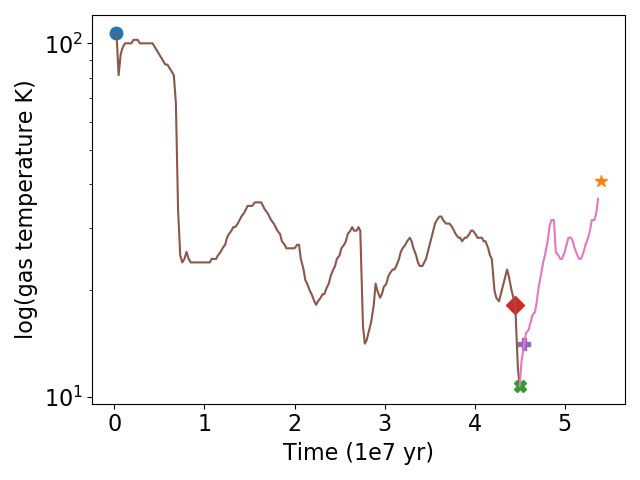}
 \includegraphics[width=1\linewidth]{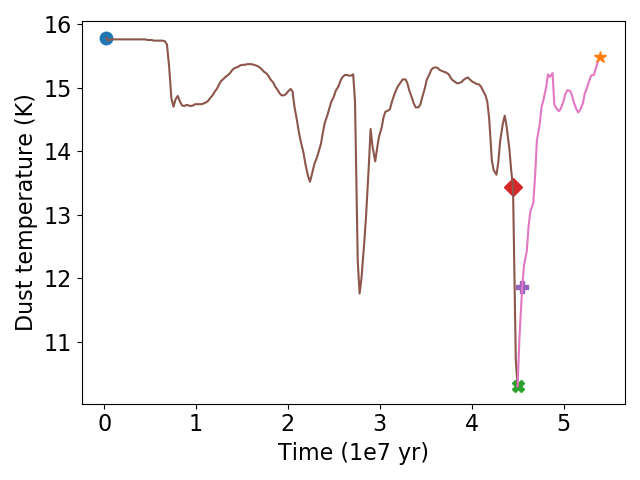}
   \caption{Gas and dust temperature as a function of time for trajectory B.}
\end{figure}


\label{lastpage}
\end{document}